%% file: jcp_bcmw.tex
\documentclass[%
 reprint,
 superscriptaddress,
 showkeys,
 amsmath,amssymb,
aps,
graphicx,
floatfix,
]{revtex4-2}

\newcommand{\eff}{\ensuremath{\mbox{\tiny eff}}}

\usepackage{xcolor}

\usepackage[version=4]{mhchem}
\usepackage{xcolor}
\usepackage{overpic}
\usepackage{booktabs}
 \usepackage{subcaption} 
\usepackage{soul}
\usepackage[normalem]{ulem}

\newcommand{\meancharge}{\ensuremath{\mathcal{Q}} }
\newcommand{\chargedeviation}{\ensuremath{\mathcal{S}} }

\usepackage{float}
\usepackage{siunitx}
\usepackage{dcolumn}
\usepackage{bm}
\usepackage{hyperref}
\usepackage{cleveref}
\usepackage{xcolor}

\graphicspath{{./images/}}

\begin{document}


\title{Counterion-controlled phase equilibria in a  charge-regulated polymer solution}



\author{Giulia L Celora}
\email[]{g.celora@ucl.ac.uk}
\affiliation{University College London, Department of Mathematics, 25 Gordon Street, London, WC1H 0AY, UK}

\author{Ralf Blossey}
\email[]{ralf.blossey@univ-lille.fr}
\affiliation{University of Lille, Unité de Glycobiologie Structurale et Fonctionnelle (UGSF), CNRS UMR8576, F-59000 Lille, France}

\author{Andreas M\"unch}
\email[]{muench@maths.ox.ac.uk}
\affiliation{Mathematical Institute, University of Oxford, Andrew Wiles Building, Woodstock Road, Oxford OX2 6GG, UK}

\author{Barbara Wagner}
\email[]{barbara.wagner@wias-berlin.de}
\affiliation{Weierstrass Institute, Mohrenstr. 39, 10117 Berlin, Germany}


\date{\today}


\begin{abstract}
We study phase equilibria in a minimal model of charge-regulated polymer solutions. Our model consists of a single polymer species whose charge state arises from protonation-deprotonation processes in the presence of a dissolved acid, whose anions serve as screening counterions. We explicitly account for variability in the polymers' 
charge states. Homogeneous equilibria in this model system are characterised by the total concentration of polymers, the concentration of counter-ions and the charge distributions of polymers which can be computed with the help of analytical approximations. 
We use these analytical results to characterise how parameter values and solution acidity influence equilibrium charge distributions and identify for which regimes uni-modal and multi-modal charge distributions arise. We then study the interplay between charge regulation, solution acidity and phase separation. 
We find that charge regulation has a significant impact on polymer solubility and allows for non-linear responses to the solution acidity: re-entrant phase behaviour is possible in response to increasing solution acidity. Moreover, we show that phase separation can yield to the coexistence of local environments characterised by different charge distributions and mixture compositions.
\end{abstract}


\maketitle 



\section{Introduction}

Solutions with charged polymers can demix into polymer-rich phases, also known as condensates. When the condensed phase remains liquid, the process yielding to demixing is known as liquid-liquid phase separation or coacervation. In recent years, the understanding of liquid-liquid phase separation (LLPS) has gained enormous interest because of its putative role in the assembly of macromolecules (mostly proteins and nucleic acids) into membrane-less organelles (also known as biomolecular condensates) in cells~\cite{Riback2020,Villegas2022}. While polymer physics theories have elucidated several aspects of phase separation in solution, it is not yet fully understood how different molecular mechanisms affect the formation, regulation and properties of biomolecular condensates in cells \cite{Villegas2022}. Challenges relate to the complexity of proteins, that are large heteropolymeric polyelectrolytes, and of the cellular environment which is maintained out of equilibrium and can itself modulate proteins properties and coacervation~\cite{Villegas2022}.

Grounded in the seminal work by Flory and Huggins (FH) on phase separation in polymer solutions, the balance between enthalpic and entropic interactions is considered to be the driving force of LLPS. Based on the simplifying assumption of polymers consisting on chemically identical units, Flory and Huggins derived a mean-field model for phase separation in two-components mixtures. Such a model has proven a useful phenomenological model also to study phase-separation in protein solutions. However, its has limited predictive power, as it misses details on the nature of the intermolecular interactions that contribute to the enthalpic part of the free energy \cite{choi_physical_2020,Villegas2022}. 

A feature common to proteins is the presence of ionizable groups, that contribute to the electrostatic interactions between proteins~\cite{shapiro_protein_2021}. Models of polyelectrolyte coacervation are commonly employed to study the role of electrostatic interactions as well as salt in LLPS.
The early key paper in the field of polyelectrolyte complexation (also called complex coacervation) remains the work by Voorn and Overbeck from 1957 \cite{VO_theory}.  
Extensions of these classical theories that capture the sequence-dependence of LLPS driven by proteins with intrinsically disordered domains, as first demonstrated in \cite{nott2015phase}, 
have employed mean-field theories of polyampholytes as underlying models of proteins. They include the Random Phase Approximation \cite{Lin2016,Meca2023}, as well as Field Theoretic Simulations \cite{zhang2020proline} for the residue specific electrostatic interactions;
recent reviews in the modern context are \cite{Brangwynne2015,Sing20172,Sing2020,rumyantsev2021polyelectrolyte}.

A limitation of all these approaches is that they assume the charge state on the polymers, such as polyelectrolytes or polyampholytes,  to be fixed; in contrast, as shown earlier on by the work of Linderstr\o{}m-Lang \cite{englander_hydrogen_1997}, the charge state of proteins is in fact regulated by the local environment, such as pH conditions, as well as by interactions between ionizable groups themselves \cite{pace_protein_2009,tanford_theory_1957}.
A key process in this context is charge regulation of the polymers or, more generally, chargeable macromolecules in the cellular context \cite{Avni2019,Avni2020}.
The charge regulation process is best explained in its most elementary variant which consists in the binding and unbinding of protons, \ce{H}$^+$, from the water solvent. It is immediately clear that this protonation-deprotonation process goes in hand-in-hand with the change of solution pH \cite{Adame-Arana2020}. More involved charge regulation processes are obviously present, e.g. in the binding of dissolved salts in solutions. 
The effect of charge regulation processes has on phase equilibria has been addressed in several recent papers \cite{Muthukumar2010,Jing2012,Salehi2016,da2018protein, Nap2016,Zheng2021,Yekymov2023}. 
However, even in simple model systems, the complexity of the interactions yields phase behaviours in multi-parameter spaces which are non-trivial to analyse. This is particularly true due to the highly non-linear free energy terms associated with electrostatic correlation effects, a key feature of liquid-liquid phase separating systems and of fundamental relevance in cell biology 
\cite{YanLevin_2002,Qin2016,Shen2017,Friedowitz2018,zhang_polyelectrolyte_2018,Zhang2021,Kumari2022}. 

In cell biology, the relation between the phase diagram on the one hand and the charge states of the macromolecules on the other \cite{Fossat2021} is of particular interest. In this paper, we address this issue on the basis of a `minimal' model which has essentially two ingredients: a basic formulation of the Voorn-Overbeek theory and the charge regulation mechanism, for which we keep track of the charge state on the polymers following the charge distribution approach developed in \cite{Avni2020}. Another key novelty that distinguishes our work from previous studies on phase separation and charge regulation processes~\cite{Muthukumar2010,Zheng2021} is that we consider the protonation-deprotonation equilibria in solution in the presence of a dissociated acid. The concentration of the counterions due to acid dissociation will turn out to be a key control parameter
in our model system. In this way, we are capable to gain insights into the coupling between charge regulation, acidity and phase separation, by linking topological changes in the coexistence curves as well as the related changes in the charge distributions on the polymers.

Our paper is organised as follows. In~\Cref{sec:model} we introduce our model for the polymer-solvent mixture. Section III covers the results we have obtained from its analysis. Section III A describes its homogeneous equilibrium states, with a focus on how the composition of the mixture affects the polymer charge. Section III B then discusses phase equilibria in our system. Finally, in Section III C we show how phase separation process itself regulates the charge state of the polymers by controlling the local environmental conditions -- here acidity. Section IV concludes and provides an outlook to further studies; in particular, we discuss
the putative relevance of our results for LLPS in biological systems. Section V contains the Appendices in which the technical results employed in the paper are derived.


\section{A model for a polymer-solvent mixture}
\label{sec:model}


\begin{figure*}[t]
\centering
\begin{subfigure}{\textwidth}
    \includegraphics[width=\textwidth]{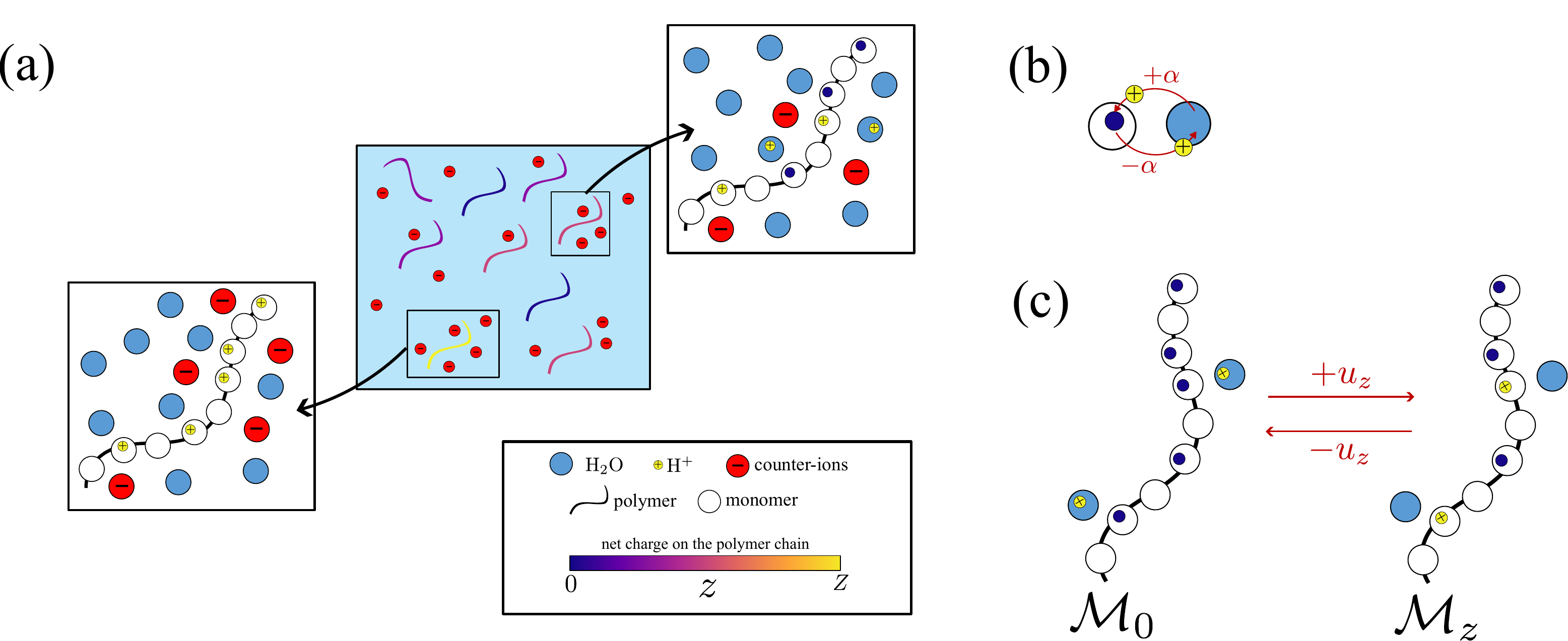}
     \captionlistentry{}
    \label{fig:1a}
\end{subfigure}
\begin{subfigure}{0.0\textwidth}
 \captionlistentry{}
    \label{fig:1b}
\end{subfigure}
\begin{subfigure}{0.0\textwidth}
 \captionlistentry{}
    \label{fig:1c}
\end{subfigure}
\caption{{\bf Mixture components.} Schematic representation of the mixture components: water (\ce{H_2}O), positive ions (\ce{H^+}), counterions and polymers chains which consists of $N$ monomers. Of the $N$ monomers polymers are made of, $Z$ have a binding site for \ce{H}$^+$ ions. The binding sites can either exist in a charged or uncharged state; as a result, polymers in the mixture can be in any charge state $z\in\left\{0,\ldots,Z\right\}$. Schematic illustrating charge regulation mechanisms: (b) for an individual monomer; (c) for an entire polymer chain.}
\label{fig:1}
\end{figure*}
 

{\bf Components of the mixture.}
The building blocks of our model and the charge regulation mechanism it entails are illustrated in Figure~\ref{fig:1}, respectively.
We consider chargeable polymers solvated in water, \ce{H_2O}, and a strong acid; here as an example, we consider hydrochloric acid, \ce{HCl}. 
Therefore, in solution, we encounter the dissociated ionic species: \ce{Cl^-} and hydronium ions \ce{H_3O^+}. 
The polymers are considered as monodisperse with $N\gg1$ monomers, of which only a subset of $Z$ monomers carries a 
protonation site, which can either be positively-charged (bound state) or neutral (unbound state). We assume that \ce{H_3O^+}, \ce{Cl^-} and the monomers making up the polymer have the same molecular volume as water, $\nu$, so that the polymers have the molecular volume $\nu_M=N\nu\gg \nu$.
As in~\cite{Avni2020}, we assume that polymers with different charge states, $z\in\left\{0,\ldots,Z\right\}$, coexist in the mixture; as a result, we have effectively $Z+1$ different polymer species in solution. Together with water, chloride and hydronium ions this gives a total of $Z+4$ species that we take into account in our mixture. For each species, we denote by $\phi_\omega$ the volume fraction, with $\omega = (s,+,\ell,z)$ = (solvent, hydronium ions, chloride ions, charged polymer). The volume fractions must satisfy a no-void condition, which guarantees that at any location space is fully occupied by the mixture:
\begin{align} \label{eq_intro:no-void}
     \phi_s+\phi_++\phi_\ell + \phi_M =1, 
\end{align} 
where
\begin{equation}\label{defphiP}
\phi_M = \sum\limits_{z=0}^{Z} \phi_z.
\end{equation}%
Furthermore, we assume that our solution is electroneutral so that
the net charge density of the mixture has to be zero, 
\begin{align}
    \phi_+-\phi_\ell+
\sum_{z=1}^{Z} \frac{z}{N} \phi_z =0. 
\label{eq_intro:electroneutrality}
\end{align}\label{eq:physical_constraints}

{\bf The free energy density of a homogeneous mixture.}
\label{sec:free energy}
We assume that the mixture is incompressible and kept at a constant temperature $T$, and describe it by a Helmholtz free energy density $f$ which consists of three contributions, similar to \cite{Zheng2021}, 
\begin{align}\label{psisum}
    f=f_1+f_2+f_3\, .
\end{align}
The chemical potentials of the different species in the mixture are then expressed in terms of derivatives 
of the Helmholtz free energy density $f$ with respect to $\phi_\omega$; these conditions are given in detail in Appendix A.
The first contribution $f_1$ in \eqref{psisum} is the standard Flory-Huggins free energy capturing the entropic contributions and an
interaction term of water and the solvated polymer
\begin{align}
\frac{f_1\nu}{k_BT}= \sum_{k\in\left\{+,\ell,s\right\}} \phi_k\ln(\phi_k)+\sum_{z=0}^{Z} \frac{\phi_z}{N} \ln(\phi_z)
+ \chi \sum_{z=0}^{Z}\phi_s \phi_{z}.\label{eq_free_energy2}
\end{align}
For simplicity, we assume the interaction parameter $\chi$ to be independent of the charge on the polymers. The second contribution, $f_2$, in \eqref{psisum} is due to charge regulation and given by
\begin{align}
\quad \frac{f_2}{k_BT}=\frac{1}{\nu_M} \sum_{z=0}^{Z} u_z\phi_z.
    \end{align}
where $u_z$ is the difference in the internal free energy (non-dimensionalised by $k_BT$) of a polymer with charge $z$ and a neutral one. By neglecting chain connectivity of the 
polymers, we can see the charged polymer 
as a mixture of an uncharged polymer and $z$ positive fixed charges. Following~\cite{Avni2020}, we specify $u_z$ as
\begin{align}
\begin{aligned}
u_z=\alpha z
+\frac{\eta z^2}{2Z}- \ln\left[\binom{Z}{z}\right].
\end{aligned}\label{eq:u_z}
\end{align}
In~\eqref{eq:u_z}, the first contribution represents the energy gain (again non-dimensionalised by $k_BT$) from occupying an additional site on the polymer 
by an $\ce{H^+}$ ion. The second term represents an additional contribution from short-range interactions between occupied binding sites whose strength is controlled by the parameter $\eta$. Finally, we have
to include the internal entropy to account for the different ways to arrange fixed charges on the binding sites.

The last term $f_3$ is the Debye-H\"uckel term, similar to  \cite{Zheng2021}, which like our reasoning for $f_2$ assumes that the 
charges on the monomers of the polymers can be treated as free ions, 
\begin{align}
    \frac{f_3}{k_BT}&=-\frac{1}{4\pi\nu} \left(\ln(1+\kappa)-\kappa +\frac{\kappa^2}{2}\right),
    \label{DH}
\end{align}
where
\begin{equation}
    \kappa^2= \lambda  \left(\phi_++\phi_{\ell}
+\frac{\nu}{\nu_M}\sum_{z=1}^{Z} z \phi_z \right)
\notag =  2\lambda \phi_\ell.
\label{eq:kappa}
\end{equation} 
Note that the term $\kappa^2$ depends on the sum of all charged molecules multiplied by their valency (as in Eq.~(6) in \cite{Zheng2021}). 
The simplified expression Eq.~(\ref{eq:kappa}) is obtained by applying 
\eqref{eq_intro:electroneutrality}. 
In Eq.~(\ref{eq:kappa}) the parameter $\lambda=4\pi \ell_B/a_w$, where $\ell_B$ is the Bjerrum length in water and $a_w=\nu^{1/3}$ 
is the size of the species in the solution. More realistic models that include polymer connectivity have, e.g., been discussed in \cite{Qin2016}. However, these include information on the specific location of the charges along the polymer chains.

{\bf The charge regulation process.}
As mentioned in the introduction, models of polymer coacervation commonly assume the charge state on the polymer phase to be fixed. In our framework, this corresponds to assuming that all protonation sites on the polymer are occupied, \emph{i.e.}, imposing in~Equations~(\ref{eq_free_energy2})—(\ref{eq:kappa}) $\phi_z=0$ for all $z=\left\{0,\ldots,Z-1\right\}$. We instead assume that charges can reversibly bind to protonation sites according to the reaction
\[ 
\ce{\mathcal{M}_{z} + H3O^+  <=> \mathcal{M}_{z+1} + H2O}, \hspace{7mm} 0 \leq z \leq Z -1, 
\]
where $\mathcal{M}_z$ represents the polymer with $z$ charges. Then, the charge states of polymers in solution is determined by imposing chemical equilibrium, instead of being prescribed \emph{a priori}. 

Making use of the definition of the Helmholtz free energy (see~\Cref{app:mu})
we have that the change in the free energy for each chemical reaction (\ce{\mathcal{M}_{z-1} + H3O^+  <=> \mathcal{M}_z + H2O}) occurring in the mixture is given by
\begin{eqnarray}
   F(T,V,N_{s} + 1,N_+ -1,\ldots, 
   N_{z-1}-1,N_z+1,\ldots) -  \nonumber \\ 
     F(T,V,N_{s},N_+,\ldots, N_{z-1},N_z,\ldots) \nonumber\\
     = \mu_s + \mu_{z} - \mu_{+}-\mu_{z-1}, 
    \quad 0 < z \leq Z. \nonumber \\
    \label{eq_app:change_free_energy}  
\end{eqnarray}

At chemical equilibrium, Eq.~\eqref{eq_app:change_free_energy} must be zero -- \emph{i.e.}, the difference in chemical potential of products and reactant of each chemical reaction must be zero. Manipulating Eq.~\eqref{eq_app:change_free_energy} we can express $\mu_z$ in terms of the chemical potential of the counterions, solvent and uncharged polymers:
\begin{equation} 
    \mu_{z}= \mu_{z-1}+\mu_+-\mu_s, \quad z = 1,\ldots, Z.
    \label{app_eq_chem_costraint}
\end{equation}
~\Cref{app_eq_chem_costraint} can be viewed as an iterative discrete map that, given $\mu_0$, defines the chemical potential of all charged polymers in terms of $\mu_+$ 
and $\mu_s$, 
\begin{equation}\label{eq:chemical_equilibrium}
        \mu_z = \mu_0+z(\mu_+-\mu_s), \quad z=1,\ldots, Z\, .
\end{equation}

Using the explicit form of the chemical potential \eqref{cpe} in \eqref{eq:chemical_equilibrium} we arrive at
\begin{eqnarray}
 u_z +\ln(\phi_z) & = & -z \chi \phi_M
+  \ln \phi_0 +z \ln\left(\frac{\phi_+}{\phi_s}\right), \nonumber \\ 
&&  z=0,\ldots,Z, \label{eq:sectionC_eq1} 
\end{eqnarray}
where $u_z$ is defined by~\Cref{eq:u_z} and  
\begin{equation}
\phi_0 =\phi_M-\sum_{z=1}^Z \phi_z.
\end{equation}

Taking the exponential of both sides of~\eqref{eq:sectionC_eq1}, we obtain a system of $Z+1$ linear algebraic equations for the volume fractions $\phi_z$; this can be solved explicitly to obtain an expression for $\phi_z$, $ z=0, \ldots, Z,$
\begin{equation}
\phi_z=\phi_M \pi_z \label{eq:charge_distribution}
\end{equation}
with
\begin{equation}
\pi_z = \mathcal{A} e^{-u_z+(\ln\phi_+-\ln\phi_s-\chi\phi_M)z}\,
\label{eq:sectionC_eq2}
\end{equation}
where
\begin{equation}
\mathcal{A}^{-1}=\sum\limits_{z=0}^{Z} e^{-u_z+(\ln\phi_+-\ln\phi_s-\chi\phi_M)z}.\label{eq:Gamma}
\end{equation}
The terms $\pi_z$ indicate the fraction of the total number of polymers in the charged state $z$ as a function of the mixture composition. By definition, their sum must be unity, $\sum_{z=0}^Z\pi_z=1$. 
Inspecting~\eqref{eq:sectionC_eq2}, we find that $\pi_z$ can be rewritten in terms of an effective charge regulation free energy, 
\begin{equation}
\pi_z=\mathcal{A}\exp\left(-u^{\mbox{\tiny eff}}_z\right),
\end{equation}
where 
\begin{subequations}
\begin{align}
    u_z^{\mbox{\tiny eff}} &= \alpha_{\mbox{\tiny eff}}z +\frac{\eta z^2}{2Z} - \ln\left[\binom{Z}{z}\right],\label{eq:effective_CR_function}\\
    \nonumber \\
\mbox{with } 
  \alpha_{\mbox{\tiny eff}}&=\alpha+\ln(\phi_s)-\ln(\phi_+)+\chi \phi_M\label{eq:alpha_eff}\, .
\end{align}%
\end{subequations}
The comparison of Eq.~\eqref{eq:effective_CR_function} to the definition of $u_z$ (see Eq.~\eqref{eq:u_z}), 
shows that, in our system, the local composition of the mixture affects the charge regulation process by controlling the energy associated with the protonation/deprotonation of a single binding site. Note the introduction of an effective parameter $\alpha_{\mbox{\tiny eff}}$ that includes a composition-dependent correction to the `bare' linear term in $u_z$.
As in \cite{Avni2020}, we find that the ion concentration in solution, $\phi_+/\phi_s$, affects the effective binding energy. Furthermore, by introducing the Flory-Huggins term in the free-energy, we have that the polymer concentration, $\phi_M$, itself affects the binding of ions in solution (see the last term in Eq.~(\ref{eq:alpha_eff})).

Using~\eqref{eq:sectionC_eq2} to eliminate $\phi_z$ ($z=0,\ldots,Z$) from the definition of free energy density (see~\eqref{psisum}-\eqref{DH}) we obtain
the expression for the free energy for an ionic solution with charge regulating polymers
\begin{eqnarray}
\frac{\nu f_{\mbox{\tiny CR}}}{k_BT} & = & \phi_+\ln \left[\phi_+\right]+\phi_\ell \ln\left[\phi_\ell\right]+\phi_s\ln[\phi_s]+\chi \phi_M\phi_s \nonumber \\
&+& \frac{\phi_M\nu}{\nu_M}\left(\ln\phi_M + 
\ln \left[\mathcal{A}\right]\right) \nonumber \\
&+& \frac{\mathcal{Q}\phi_M\nu}{\nu_M}\left(\ln\phi_+-\ln\phi_s-\chi \phi_M\right) \nonumber \\
&-& \frac{1}{4\pi}\left(\ln(1+\kappa)
+\frac{\kappa(\kappa-2)}{2} \right) 
\label{eq:reformulated_free_energy}
\end{eqnarray}%
where $\kappa=\sqrt{2\lambda\phi_\ell}$ and we have introduced the variable $\mathcal{Q}$ that represents the mean charge of the polymer phase
$\mathcal{Q}=\sum_{z=0}^{Z}z\pi_z$
While we have defined the free energy in terms of the variables $\phi_s$, $\phi_+$,  $\phi_M$ and $\phi_\ell$, the degrees of freedom of the model 
can be reduced to only two by observing the two constraints (no-void and electro-neutrality) formulated in \eqref{eq_intro:no-void} and \eqref{eq_intro:electroneutrality}, that is 
$\phi_s =1-\phi_M-\phi_\ell-\phi_+$,
and 
$   \phi_+
= -\frac{\phi_M}{N}\mathcal{Q}+\phi_\ell $.

These determine $\phi_s$ and $\phi_+$ in terms of $\phi_M$ and $\phi_\ell$, albeit, in the case of $\phi_+$, only implicitly.

\section{Results}
\label{sec:results}

In the current work, we focus on the interplay between charge regulation processes and phase separation. Our analysis highlights the key role of parameter $\eta$ in the equilibrium properties of the system. We, therefore, consider it as a free parameter while fixing the others. Based on previous works, we set $\lambda=26.68$~\cite{Zheng2021} and $\nu\approx3.1 \times 10^{-23} \hbox{ ml}$~\cite{Zheng2021}. The number of monomers in the protein is set to $N = 100$; of these, we assume that $Z=20$ have a \ce{H+} binding site. We set $\alpha=-6.5$ so that it is energetically favourable for an individual binding site to be occupied (see~\Cref{fig:1b}). The temperature is fixed to $T=298$ K and the Flory parameter to $\chi=0.95$; the latter value is chosen so that phase separation is observed -- even when considering a neutral polymer (see~\Cref{sec:phase_separation_fixed_charges}).

\begin{figure*}[ht!]
\includegraphics[width=\textwidth]{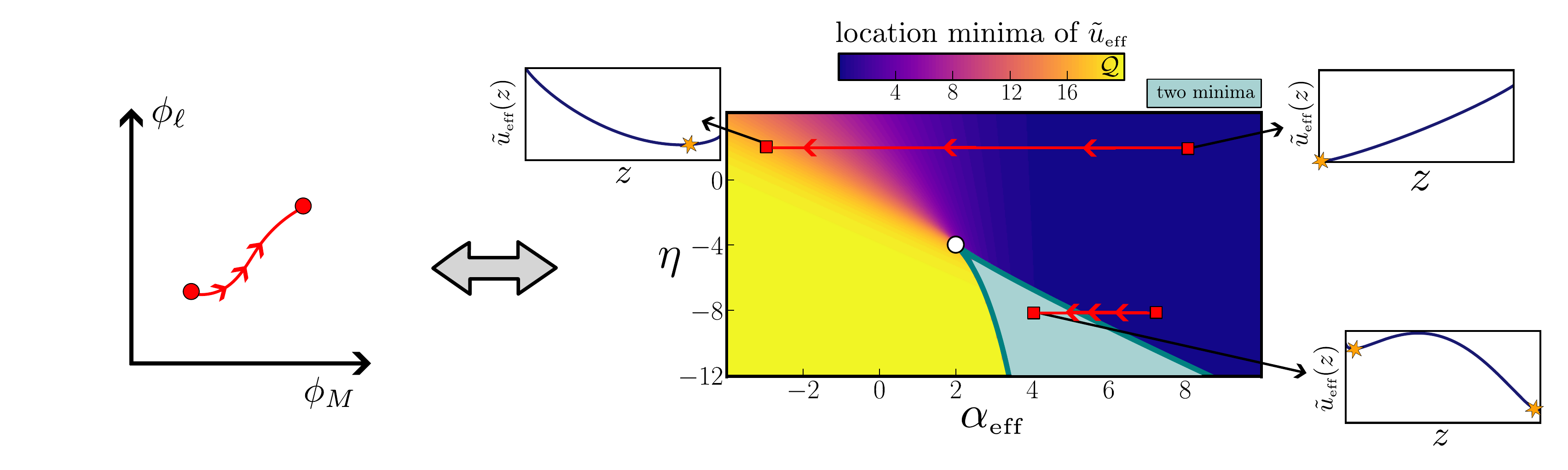}
\caption{{\bf Composition-dependent charged states.} Parameter diagram for the charge distribution of homogeneous states as a function of $\alpha_{\mbox{\tiny eff}}$ and $\eta$, obtained by computing the extrema of $\tilde{u}_{\mbox{\tiny eff}}$ (see~\eqref{eq:u_eff}). The insets show $\tilde{u}_{\mbox{\tiny eff}}$ for specific values of $\alpha_{\mbox{\tiny eff}}$ and $\eta$. In the green region $\tilde{u}_{\mbox{\tiny eff}}$ has two minima; outside this region a unique minimum exists and its position is indicated by the colorbar above the diagram. 
The change of the effective binding energy parameter $\alpha_{\mbox{\tiny eff}}$ (red path in the ($\phi_M$,$\phi_\ell$)-plane on the left corresponds to moving along a horizontal line in the ($\alpha_{\mbox{\tiny eff}}$,$\eta$)-plane).}
\label{fig:2}
\end{figure*}

\subsection{Analysis of homogeneous equilibrium states.}
\label{sec:homogeneous_phase}
We first study the properties of homogeneous equilibrium states that arise in our model. We are specifically interested in how the charge distribution of the polymers, $\pi_z$, depends on the mixture composition, $\phi_M$ and $\phi_\ell$. This is obtained by solving the non-linear system of algebraic equations given by Eqs.~\eqref{eq_intro:no-void}, \eqref{eq_intro:electroneutrality} and \eqref{eq:charge_distribution}-\eqref{eq:Gamma}. Generally, this can not be done analytically and requires numerical approaches. However, we make the following observations.

1.) In the case $\eta = 0$ (i.e., independent ion adsorption), the charge distributions $\pi_z$ is binomial, which can be approximated by a Gaussian distribution in $z$ when taking the maximum charge, $Z\gg1$;

2.) For $z$ taken as a continuous variable, we can approximate the effective charge regulation free energy $u_{\mbox{\tiny eff}}$ as
\begin{equation}
    \tilde{u}_{\mbox{\tiny eff}}(z) = z\alpha_{\mbox{\tiny eff}}+\frac{z^2\eta}{2Z} +z\ln\frac{z}{Z}+(Z-z)\ln\left(1-\frac{z}{Z}\right)\label{eq:u_eff}
\end{equation}
in the limit $Z\gg1$ and $z\in(0,Z)$ (for the details, see ~\Cref{app:Gaussian approximation}). In~\Cref{fig:2}, we summarise how the number and location of the local minima of $\tilde{u}_{\eff}$ is controlled by the mixture composition -- \emph{i.e.,} the value of the parameter $\alpha_{\eff}$. When $\tilde{u}_{\eff}$ has a single minimum, then we can estimate $\pi_z$ within a saddle-point approximation that we detail in Section V B.  We find that for $\eta>-4$, we can approximate the charge distribution by a Gaussian distribution whose mean is determined by the minimum of $\tilde{u}_{\eff}$.

3) The saddle-point approximation is not always valid for $\eta<-4$.
The breakdown of the saddle-point approximation is due to the appearance of multiple extrema for the function $\tilde{u}_{\eff}$ (see green area in~\Cref{fig:2}) that is reflected in the charge distribution $\pi_z$ having multiple peaks. In this case of failure of the saddle-point approximation, we need to resort to numerical methods of computation.

This general feature of unimodality vs. multimodality of the charge distribution is summarised in Figure \ref{fig:2} which displays the $(\eta,\alpha_{\mbox{\mbox{\tiny eff}}})$ diagram. 
As shown, we can identify two characteristic regimes depending on the value of $\eta$: when $\eta>-4$, $\alpha_{\mbox{\tiny eff}}$ (\emph{i.e.}, the mixture composition), controls the location of the minimizer of $\tilde{u}_{\mbox{\tiny eff}}$ which is always unique; similarly of $u_z^{\mbox{\tiny eff}}$. 
When $\eta<-4$, $\alpha_{\mbox{\tiny eff}}$ (\emph{i.e.}, the mixture composition), controls both the location and the number of minimizers of $\tilde{u}_{\mbox{\tiny eff}}$, and likewise of $u_z^{\mbox{\tiny eff}}$. We note that transitions between unimodality to multimodality in charge regulating systems had earlier been seen in \cite{Avni2020}. 

We now discuss the three different cases of interest separately in more detail.

\subsubsection{The case $\eta = 0$: independent ion adsorption.}

By setting $\eta=0$, Eqs.~(\ref{eq:distribution_unimodal}) are exact and this can be shown without the need of any approximation. Indeed, we have that $\mathcal{A}$ can be evaluated explicitly: $\mathcal{A}=(1+ e^{-\alpha_{\mbox{\tiny eff}}})^{-Z}$.
We obtain
\begin{equation}\label{pp}
\pi_z=\binom{Z}{z} p^z \left(1-p\right)^{Z-z},
\end{equation}
where 
\begin{equation}
p=\frac{e^{-\alpha_{\mbox{\tiny eff}}}}{1+e^{-\alpha_{\mbox{\tiny eff}}}}.
\label{eq:prob_bounding_eta_0}
\end{equation}
Thus the distribution of polymer states, 
normalised by the total polymer concentration
$\phi_M$, has the form of a binomial distribution 
$ B(Z,p)$. We can explain the appearance binomial distribution of the charge state of polymers intuitively. When $\eta=0$ there is no correlation of different binding sites; thus the state of each of the $Z$ sites can be treated as an independent Bernoulli random variable with probability of success (\emph{i.e.}, binding) equal to $p$ (see Eq.~(\ref{eq:prob_bounding_eta_0})). 



\subsubsection{$\eta > -4$: the general unimodal case.}

\begin{figure*}[t]
    \begin{subfigure}{\textwidth}
    \includegraphics[width=\textwidth]{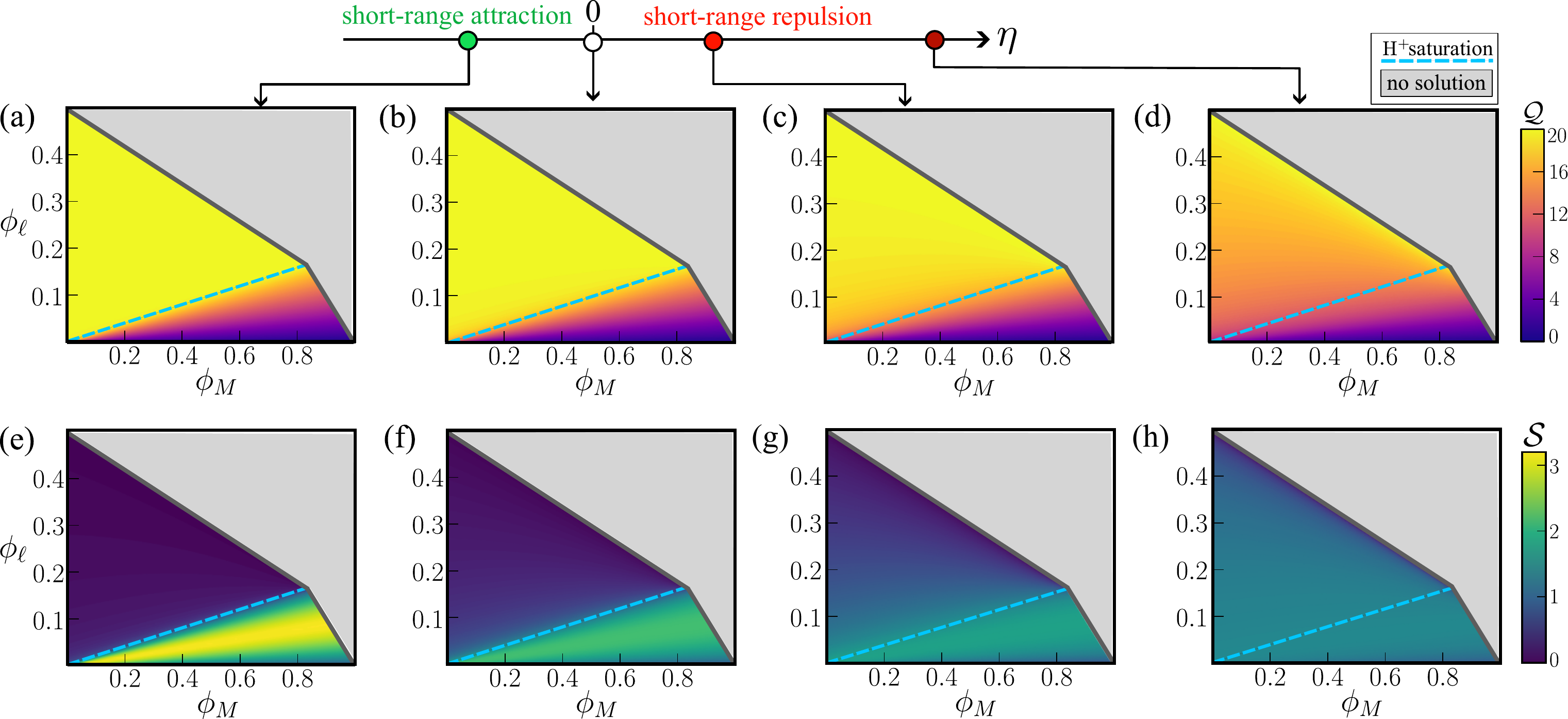}
    \captionlistentry{}
    \label{fig:3a}
    \end{subfigure}
        \begin{subfigure}{\textwidth}
    \captionlistentry{}
    \label{fig:3b}
    \end{subfigure}
            \begin{subfigure}{\textwidth}
    \captionlistentry{}
    \label{fig:3c}
    \end{subfigure}
            \begin{subfigure}{\textwidth}
    \captionlistentry{}
    \label{fig:3d}
    \end{subfigure}
            \begin{subfigure}{\textwidth}
    \captionlistentry{}
    \label{fig:3e}
    \end{subfigure}
            \begin{subfigure}{\textwidth}
    \captionlistentry{}
    \label{fig:3f}
    \end{subfigure}
            \begin{subfigure}{\textwidth}
    \captionlistentry{}
    \label{fig:3g}
    \end{subfigure}
            \begin{subfigure}{\textwidth}
    \captionlistentry{}
    \label{fig:3h}
    \end{subfigure}
    \caption{{\bf Composition dependence of mean charge and standard deviation.} (a)-(d) Series of surface plots illustrating how the mean charge, $\meancharge$, depends on the local composition of the mixture $(\phi_M,\phi_\ell)$, for different values of the parameter $\eta$ -- from left to right: $\eta=-2$; $\eta=0$; $\eta=2$ and $\eta=5$ (short-range repulsion between bounded charges). (e)-(g) Same as panels (a)-(d) but illustrating the computed standard deviation, \chargedeviation. The dotted light blue lines indicate the salt concentration at which the concentration of \ce{H+} ions in solution equilibrates the concentration of binding sites, \emph{i.e.} $\phi_\ell=Z\phi_M/N$. Other parameters are set to default values given at the start of~\Cref{sec:results}.}
    \label{fig:3}
\end{figure*}

When the value of $\alpha_{\mbox{\tiny eff}}$ is such that we lie outside the green region in~\Cref{fig:2}, the charge distribution $\pi_z$ is unimodal with most polymers having a charge state similar to $z\approx\meancharge$, defined as the unique minimizer of~\eqref{eq:u_eff}. As shown in~\Cref{app:Gaussian approximation}, $\pi_z$ can be approximated by a Gaussian whose mean charge \meancharge and standard deviation \chargedeviation, can be written as
\begin{subequations}
    \begin{align}
        \meancharge&= Z p,\quad \chargedeviation^2 = \frac{Z p(1-p)}{\eta p (1-p) + 1},
    \end{align}\label{eq_eta_non_zero:moments}%
    where $p\in(0,1)$ is implicitly defined by
    \begin{align}
       p=\frac{e^{-\alpha_{\mbox{\tiny eff}}-p\eta}}{1+e^{-\alpha_{\mbox{\tiny eff}}-p\eta}}.\label{eq_eta_non_zero:implicit_p}
    \end{align}\label{eq:distribution_unimodal}%
    \end{subequations}
In the case $\eta>-4$, $\chargedeviation^2$ is guaranteed to be positive independently of the value of $p\in(0,1)$.
When comparing the exact form of $\mathcal{Q}$ and $\mathcal{S}$ in the case $\eta=0$ (see~\Cref{pp}) and the approximated form for $\eta\neq 0$ (see~\Cref{eq_eta_non_zero:moments}), we find clear parallelisms. When considering $\eta\neq0$, the model captures the extra energy contributions due to the interaction of the charges on the polymers. Unlike from the case $\eta=0$, this introduces correlation amongst the state of binding sites (occupied or unoccupied) on the same polymer. Nonetheless, we may still interpret $p$ in ~\Cref{eq_eta_non_zero:implicit_p} as the binding probability for an \ce{H+} ion to a free binding site. We note that the analogy with the binomial distribution is not exact and difference emerges when comparing the second moments -- here the variance $\chargedeviation^2$ -- which explicitly depends on $\eta$. 
When considering states with the same mean charge $\meancharge$, we have that $\eta>0$ (short-range repulsion) results in a reduction of the variance of the distribution. In contrast, negative values of $\eta$ yield to wider distributions, \emph{i.e.}, larger values of $\mathcal{S}$. 
So far, we have considered $\alpha_{\mbox{\tiny eff}}$ as a prescribed parameter. However, as illustrated in~\Cref{eq:distribution_unimodal}, $\alpha_{\mbox{\tiny eff}}$ is determined by the mixture composition -- \emph{i.e.}, the values of $\phi_M$ and $\phi_\ell$. The computation of the corresponding concentration diagrams requires solving highly non-linear equations, for which existence and uniqueness of solution may not be guaranteed. Due to the physical constraints in the system (no-void and electro-neutrality), homogeneous equilibrium states only exists when $\phi_M$ and $\phi_\ell$ satisfy:
\begin{subequations}
    \begin{align}
        1-\phi_M-\phi_\ell >0,\label{cond1:m}\\
        1+\phi_M \left(\frac{Z}{N}-1\right) - 2\phi_\ell>0,\label{cond2:m}   \end{align}\label{cond_existence_m}%
\end{subequations}
We can prove that such homogeneous states are unique (see Appendix~\ref{app:derivation unimodal} for details). We obtain the solutions numerically via Newton's method and use the approximation to estimate how $\mathcal{Q}$ and $\mathcal{S}$ vary as a function of the mixture composition. Results for different values of $\eta>-4$ are shown in \Cref{fig:3}.

When $\eta$ is negative (as in~\Cref{fig:3a}), the fully-charged state is the most energetically favourable for the polymers -- recall $\alpha$ is also taken to be negative. As a result, whenever the concentration of \ce{H+}--ions in the mixture exceeds the concentration of the binding sites (\emph{i.e.}, $\phi_\ell>(Z/N)\phi_M$ -- above the dotted light-blue curve in~\Cref{fig:3}), the polymers will be in a fully-charged state-- as \meancharge attains its maximum value (see panel (a)) while \chargedeviation its minimum (see panel (b)). In contrast, when the concentration of \ce{H+}--ions in the mixture is lower than the concentration of the binding sites (\emph{i.e.}, $\phi_\ell<(Z/N)\phi_M$), the charges are on average distributed homogeneously between the polymers, $\mathcal{Q}\approx N\phi_\ell/(Z\phi_M)$. This can be shown systematically, by considering the limit $\alpha\to -\infty$ when estimating $p$ (results not shown). As $\eta$ increases it becomes less energetically favourable for \ce{H+}--ions to bind to the polymers that tend to remain in a less charged state even when $\phi_\ell>(Z/N)\phi_M$. As expected, we find that the largest value of $\mathcal{S}$ decreases with $\eta$. However, when considering the impact of $\eta$ on $\mathcal{S}$ for a specific mixture composition, there is no general trend. For ion-saturated mixture compositions, $\mathcal{S}$ increases with $\eta$, while for ion-limiting mixture compositions, $\mathcal{S}$ decreases with $\eta$.

\subsubsection{Multi-modal charge distributions: charge demixing.} 

We now investigate the equilibrium charge distribution for values of $\eta<-4$. As discussed at the beginning of this section, in this regime, the saddle-point approximation breaks down and bimodal charge distributions are expected. 

We compute the full charge distribution, $\left\{\pi_z\right\}$, solving the non-linear algebraic system given by Eqs.~\eqref{eq_intro:no-void}, \eqref{eq_intro:electroneutrality} and \eqref{eq:charge_distribution}-\eqref{eq:Gamma} using Newton's method with arc-length continuation (used to find good initial guesses for the Newton's step).
We conjecture that $\left\{\pi_z\right\}$ is still uniquely defined even when we are in regimes for which the charge distribution has multiple peaks (\emph{i.e.}, when we enter the green area in~\Cref{fig:2}); this is strongly supported by our numerical investigation but an analytical proof of the result is beyond the scope of this work.

\begin{figure*}[htb]
    \begin{subfigure}{\textwidth}
    \includegraphics[width=0.75\textwidth]{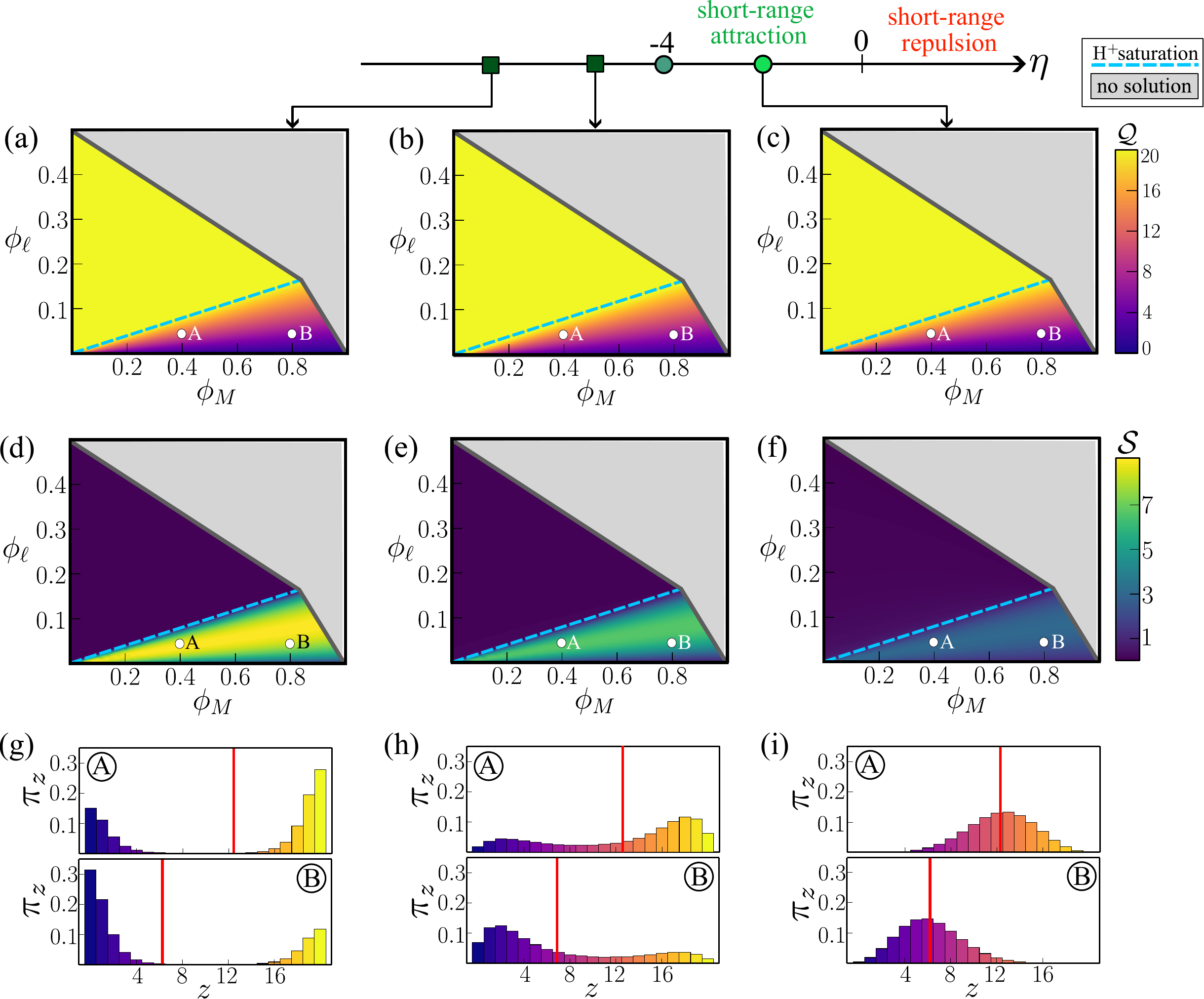}
    \captionlistentry{}
    \label{fig:4a}
    \end{subfigure}
      \begin{subfigure}{\textwidth}
    \captionlistentry{}
    \label{fig:4b}
    \end{subfigure}
          \begin{subfigure}{\textwidth}
    \captionlistentry{}
    \label{fig:4c}
    \end{subfigure}
          \begin{subfigure}{\textwidth}
    \captionlistentry{}
    \label{fig:4d}
    \end{subfigure}
          \begin{subfigure}{\textwidth}
    \captionlistentry{}
    \label{fig:4e}
    \end{subfigure}
          \begin{subfigure}{\textwidth}
    \captionlistentry{}
    \label{fig:4f}
    \end{subfigure}
          \begin{subfigure}{\textwidth}
    \captionlistentry{}
    \label{fig:4g}
    \end{subfigure}
          \begin{subfigure}{\textwidth}
    \captionlistentry{}
    \label{fig:4h}
    \end{subfigure}
          \begin{subfigure}{\textwidth}
    \captionlistentry{}
    \label{fig:4i}
    \end{subfigure}
    \caption{{\bf Composition dependence of the mean charge and standard deviation for  $\eta < 0$.}
    Series of surface plots illustrating how (a)-(c) $\meancharge$ and (d)-(f), depends on the local composition of the mixture $(\phi_M,\phi_\ell)$, for different values of the parameter $\eta$: (left column) $\eta=-7$; (middle column) $\eta=-5$; and (right column) $\eta=-2$ (same as~\Cref{fig:3a}). The dotted light blue lines indicate the salt concentration at which the concentration of \ce{H+} ions in solution equilibrates the concentration of binding sites, \emph{i.e.} $\phi_\ell=Z\phi_M/N$.
    (g)-(i) Plots of the charge distribution, $\pi_z$ (see \eqref{eq:charge_distribution}) for specific values of $(\phi_M,\phi_\ell)$ (see white dots in panels (a)-(f)); the red vertical lines indicate the mean of the distribution, $\mathcal{Q}$. Other parameters are set to default values given at the start of~\Cref{sec:results}. }
    \label{fig:4}
\end{figure*}

The results are shown in~\Cref{fig:4} in which we compare the homogeneous equilibrium states for $\eta=-7$ (left column), $\eta=-5$ (middle column) and $\eta=-2$ (right column). Interestingly, we find that $\mathcal{Q}$ is almost insensitive to changes in $\eta$ (recall that here $\alpha=-6.5\ll 0$); both below and above the \ce{H+}-saturation curve the mean charge is not affected by increasing of the short-range attractions between bounded charges (\emph{i.e.}, moving from right to left in~\Cref{fig:4}). In contrast, the behaviour of the standard deviation $\mathcal{S}$ changes significantly with $\eta$; particularly for mixture compositions below the saturation curve. Overall, we find that the more negative $\eta$, the larger the maximum value of $\mathcal{S}$. When $\eta<-4$ (see first and middle column in~\Cref{fig:4}), large values of the variance $\mathcal{S}$ are attained by allowing charges to be distributed unevenly between polymers -- \emph{i.e.}, $\pi_z$ has a bimodal profile (see panels (g) and (h) in~\Cref{fig:4}). For values of $\eta$ near the critical threshold $\eta=-5$ (see panel (h)), we find broad distributions, with polymers in all charge states present in the solution. In this case, the peaks in the distributions occur away from $\meancharge$ (see vertical red line) suggesting that most polymers have a charge state that deviates from the mean. As we take $\eta\ll-4$ (see panel (g)), $\pi_z$ becomes more skewed towards the extreme states, $z=0$ and $z=1$, and the large values of $\mathcal{S}$ are due to the differential partitioning of the charges rather than $\pi_z$ having a broader support. This is because the intermediate charge states, $z\approx Z/2$, become energetically unfavourable and most polymers exist either in a poorly-charged ($z\approx 0$) or in a highly-charged state ($z\approx 1$). In this regime, changes in the mixture composition only impact the relative fraction of the polymers in poorly-charged and in highly-charged states thus allowing $\meancharge$ to attain all values in the interval $[0,Z]$. From this point of view, the model could be approximated by a two-population model: either neutral or fully charged polymers that coexist under proper conditions. This is similar to the approach adopted in~\cite{Adame-Arana2020}. 

\subsection{Demixing in solutions of charged polymers}
\label{sec:coexistence_curves}

In the previous section we have discussed how charge regulation affects the homogeneous equilibrium states of the mixture. In particular, we find that the mixture composition modulates the equilibrium charge distribution. Due to the physical constraints on the volume fractions -- \emph{i.e.}, no-void and electro-neutrality -- at equilibrium the mixture composition is well-defined by the volume fraction of two species. Here we have chosen: the total volume fractions of polymers, $\phi_M$ and counterions, $\phi_\ell$. 

We now investigate how charge regulation impacts the solubility of charged polymers. The calculation of the phase diagrams follows standard procedures -- details are given in Appendix~\ref{app:ph}. We denote by $\left\{\phi^{I}_\omega\right\}$ and  $\left\{\phi^{II}_\omega\right\}$ the volume fraction of species in the dilute (\emph{i.e.}, polymer depleted) and condensed (\emph{i.e.}, polymer rich) phases, respectively. Importantly, in constructing the phase diagrams we allow the ions to be distributed asymmetrically between the dilute and condensed phases. As a result, the tie-lines (\emph{i.e.}, the curve connecting coexisting states) can have non-zero gradients. This leads to the mean electrostatic potential being different in the dilute ($\psi^I$) and condensed ($\psi^{II}$) phases. The difference $\Delta \psi=\psi^{II}-\psi^{I}$ is known as the \emph{Galvani potential} \cite{Zhang2021}. For any value of the model parameters the phase diagrams are practically computed in \texttt{Julia} using the \texttt{BifurcationKit} package \cite{veltz_bif_kit} for numerical continuation. 

As mentioned in~\Cref{sec:free energy}, most models of phase separation assume that the charges on the polymers are fixed. In order to highlight the role of charge regulation in phase separation, we first investigate demixing for a solution of polymers with a fixed charge, $Z$. While in the charge regulation (CR) model the charge distribution, $\pi_z$ is obtained by minizing the free energy $f$ (see Eqs~(\ref{psisum})-(\ref{eq:kappa})), in a fixed charge (FC) model, $\pi_z$ is prescribed via a delta function $\pi_z=\delta(z-Z)$. Substituting $\phi_z=\phi_M\delta(z-Z)$ into Eqs.~(\ref{psisum})-(\ref{eq:kappa})), we obtain the free energy for the FC model, $f_{\mbox{\tiny FC}}$, as
\begin{equation}
\begin{aligned}
    \frac{f_{\mbox{\tiny FC}}\nu}{k_BT} &=  \phi_+\ln \left[\phi_+\right]+\phi_\ell \ln\left[\phi_\ell\right]+\phi_s\ln[\phi_s]+ \frac{\phi_M}{N}\ln\phi_M\\&+ \frac{u_Z\phi_M}{N}+\chi \phi_M\phi_s - \frac{1}{4\pi}\left(\ln(1+\kappa)
+\frac{\kappa(\kappa-2)}{2} \right) 
\end{aligned}\label{eq:f_FC}
\end{equation}
where $\kappa=\sqrt{2\lambda \phi_\ell}$ (as before) and $u_Z$ is defined as in~\eqref{eq:u_z}. As before, the system must also satisfy the electro-neutral and no-void constraints (see Eqs.~\eqref{eq_intro:no-void} and \eqref{eq_intro:electroneutrality}). 

\subsubsection{Phase diagrams for macromolecules with a fixed charge.} 

\begin{figure*}[ht!]
    \begin{subfigure}{\textwidth}
    \includegraphics[width=\textwidth]{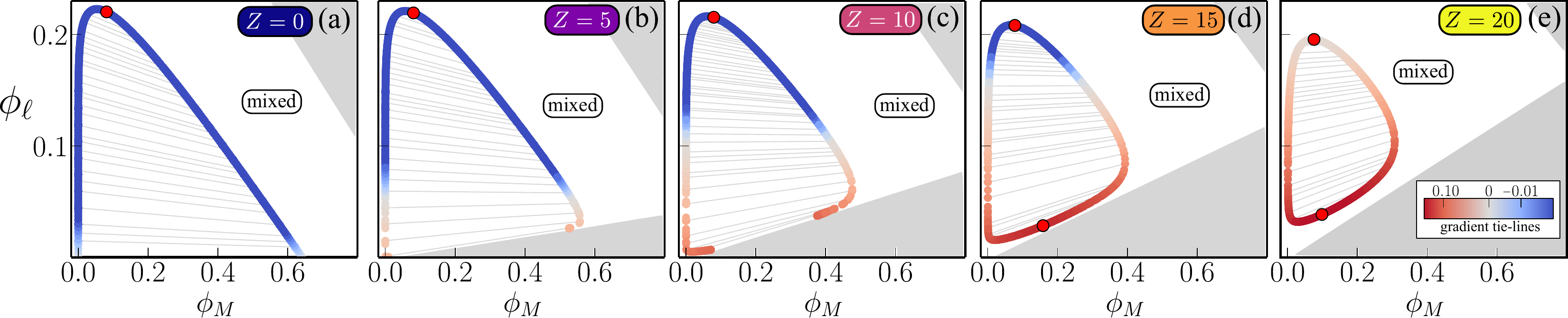}
    \captionlistentry{}
    \label{fig:4a_panelA}
    \end{subfigure}
        \begin{subfigure}{\textwidth}
    \captionlistentry{} 
    \label{fig:4a_panelB}
    \end{subfigure}
            \begin{subfigure}{\textwidth}
    \captionlistentry{}
    \label{fig:4a_panelC}
    \end{subfigure}
            \begin{subfigure}{\textwidth}
    \captionlistentry{}
    \label{fig:4a_panelD}
    \end{subfigure}
            \begin{subfigure}{\textwidth}
    \captionlistentry{}
    \label{fig:4a_panelE}
    \end{subfigure}
    \vspace{-8mm}
    \caption{{\bf Phase diagram topologies for polymers with fixed charges.} In the different panels, the following fixed charge values $Z$ have been chosen: 
    (a) $Z=0$, (b) $Z=5$, (c) $Z=10$, (d) $Z=15$ and (e) $Z=20$. The colour scale indicates the gradient of the tie-lines while tie-lines connecting coexisting states are indicated in light grey. The area of the ($\phi_M$,$\phi_\ell$) space that are unphysical for our model (\emph{i.e.}, electroneutrality is not satisfied) are shadowed in grey. Critical points at which the two coexisting phases become indistinguishable are denoted by the red circles. Other parameters are set to default values given at the start of~\Cref{sec:results}. }
    \label{fig:5}
\end{figure*}

In~\Cref{fig:5}, we present the phase diagram for increasing values of the fixed charge on the macromolecules, $Z$. In these diagrams, regions of mixing and demixing are separated by the binodal (or coexistence) curves. Along the binodal, we highlight the gradient of the tie-lines: positive gradients (in red) indicate the counterions concentration is higher in the condensed phase (II); in contrast, negative gradients (in blue) imply counterions accumulate in the dilute phase (I). We note that, besides the constraints~\eqref{cond_existence_m}, in the fixed charge model, the electroneutrality condition also requires $\phi_\ell>Z/N \phi_M$. 

Starting from the case of neutral polymers (see~\Cref{fig:4a_panelA}), we recover a coexistence curve analogous to the one obtained in previous works on coacervates \cite{Sing20172,Zhang2021}. Here the region of demixing is enclosed by a single open curve (the bimodal) and a unique critical point (highlighted in red) exists. Furthermore, the tie-lines have a negative gradient, suggesting that more counterions accumulate in the dilute instead of the condensed phase. The gradient steepens near the critical point, while tie-lines are almost horizontal when the counterions are dilute ($\phi_\ell^{(I)}\ll 1$). As we increase $Z$ the fixed charge on the polymers (see~\Cref{fig:4a_panelB}-\ref{fig:4a_panelC}), the demixing region is affected only for small values of $\phi_\ell$; this is primarily due to intersection of the bimodal curve with the boundary of the feasibility region ($\phi_\ell=Z/N\phi_M$). 
Since the latter curve has a positive gradient, this enforces the tie-lines to change their orientation as they approach the boundary of the feasibility region. If we increase the charge on the polymers even further (see~\Cref{fig:4a_panelD}-\Cref{fig:4a_panelE}), we find the demixing region shrinks and its topology changes into a closed-loop with the emergence of a second critical point. We find also a complete inversion in the slope of the tie-lines compared to the neutral case. If we were to increase $Z$ even further, the miscibility gap will disappear (results not shown). 

We conclude that overall fixed charges reduce the solubility of polymers in solution. 

\subsubsection{Phase diagrams for charge-regulating polymers.}
\label{sec:phase_separation_fixed_charges}
\begin{figure*}[htb]
    \centering 
    \vspace{5mm}
    \begin{subfigure}{\textwidth}
    \includegraphics[width=0.9\textwidth]{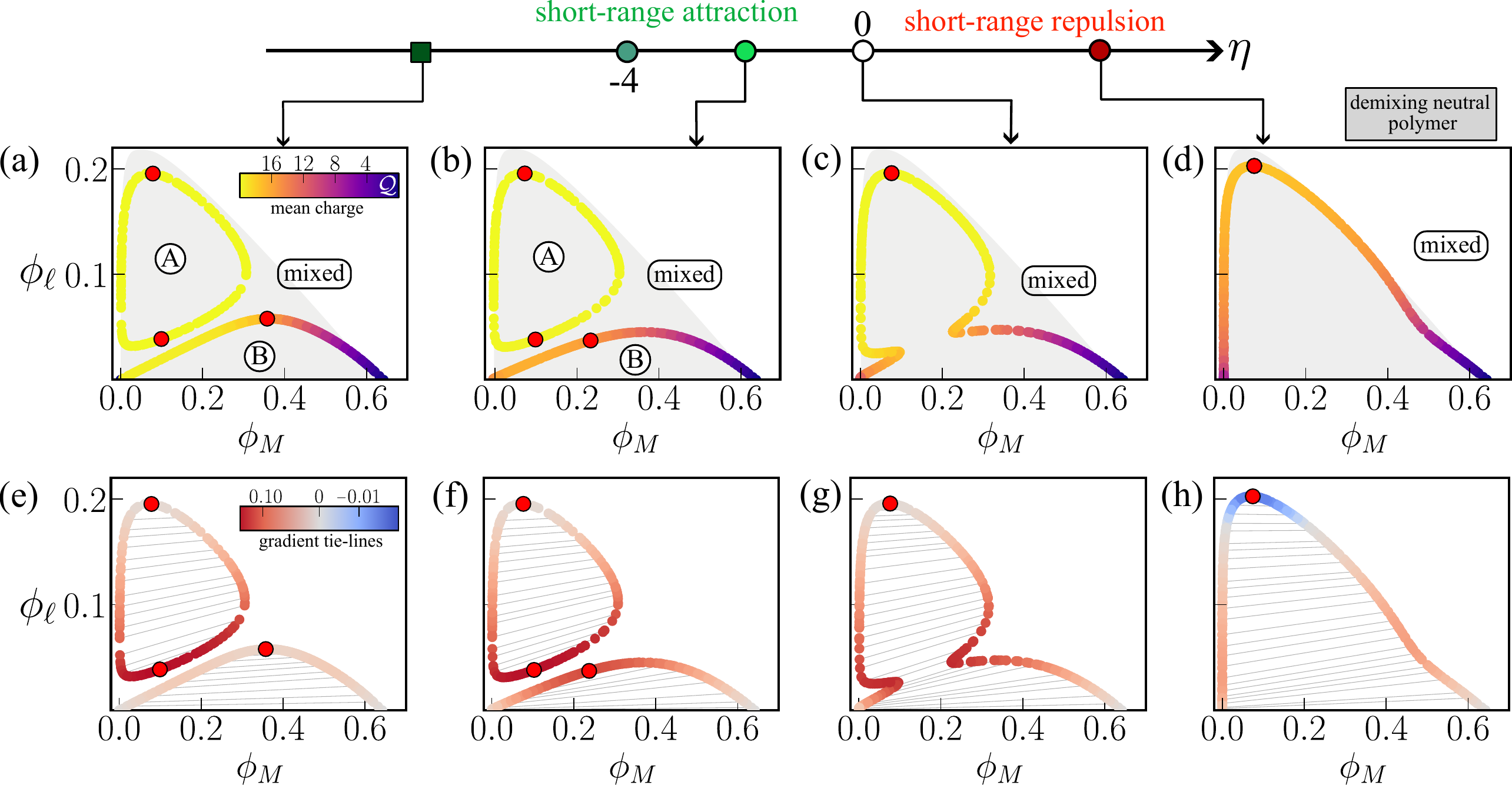}
    \captionlistentry{}
    \label{fig:5_panelA}
    \end{subfigure}
    \begin{subfigure}{0.0\textwidth}
    \captionlistentry{}
    \label{fig:5_panelB}
    \end{subfigure}
    \begin{subfigure}{0.0\textwidth}
    \captionlistentry{}
    \label{fig:5_panelC}
    \end{subfigure}
      \begin{subfigure}{0.0\textwidth}
    \captionlistentry{}
    \label{fig:5_panelD}
    \end{subfigure}
      \begin{subfigure}{0.0\textwidth}
    \captionlistentry{}
    \label{fig:5_panelE}
    \end{subfigure}
          \begin{subfigure}{0.0\textwidth}
    \captionlistentry{}
    \label{fig:5_panelF}
    \end{subfigure}
          \begin{subfigure}{0.0\textwidth}
    \captionlistentry{}
    \label{fig:5_panelG}
    \end{subfigure}
          \begin{subfigure}{0.0\textwidth}
    \captionlistentry{}
    \label{fig:5_panelH}
    \end{subfigure}
    \vspace{-5mm}
    \caption{{\bf Phase diagram topologies for different values of the charge regulation parameter $\eta$}. The following values were chosen: $\eta=-7.0$ (first column); $\eta=-2$ (second column); $\eta=0$ (third column) and $\eta=2$ (fourth column). In panels (a)-(d) the colour map indicates the mean charge $\mathcal{Q}$ along the binodal curve; the grey area indicates the demixing region for the neutral polymer solution (same as in~\Cref{fig:4a_panelA}). In panels (e)-(h), the colour map indicates the gradient of the tie-lines (indicated in light grey). Critical points at which the two coexisting phases become indistinguishable are denoted by red circles. Other parameters are set to default values given at the start of~\Cref{sec:results}. }
    \label{fig:6}
\end{figure*}

In~\Cref{fig:6}, we illustrate the characteristic topologies of the phase diagram for charge-regulating polymers for different values of $\eta$. In these diagrams, regions of mixing and demixing are separated by the binodal curves. In Figures~\ref{fig:5_panelA}-\ref{fig:5_panelD}, we depict along the bimodal the mean charge on the polymers, $\mathcal{Q}$. In Figures~\ref{fig:5_panelE}-\ref{fig:5_panelH}, we illustrate the same phase diagrams but highlight along the binodal the gradient of the tie-lines (see grey curves). Interestingly, we find that the phase diagrams can be significantly different from each other depending on the value of the charge regulation parameter $\eta$. In particular, we find that, for strong short-range repulsion between occupied binding sites, \emph{i.e.}, $\alpha+\eta$ large and negative (first and second column in~\Cref{fig:6}), the phase diagram presents two disconnected regions of demixing -- namely A and B in~\Cref{fig:5_panelA})-- which are enclosed in the demixing region obtained for neutral polymers (see shaded area in~\Cref{fig:5_panelA}). The demixing region A in~\Cref{fig:5_panelA} lies above the \ce{H+} saturation curve (see~\Cref{fig:3a} and related discussion) and the polymers effectively behave as having a fixed charge of $Z=20$. When comparing region A in~\Cref{fig:5_panelA} (or~\Cref{fig:5_panelB}) and the demixing region in~\Cref{fig:4a_panelE}, the two overlap exactly. In contrast, the demixing region B lies fully or partially below the saturation curve. The boundary of this region is delimited by coexisting phases that differ both in the local amount of polymers as well as in their charge state -- as highlighted by the variation in the mean charge $\mathcal{Q}$. The implication of these results will be investigated in~\Cref{sec:observation_CR}. As the value of $\eta$ increases (\emph{i.e.}, it is less favourable for polymers to be in a fully charged state), the two disconnected regions merge and a single demixing region persists (see~\Cref{fig:5_panelC}).
Eventually, for $\eta$ sufficiently positive, the phase diagram converges to the one of neutral polymers (see~\Cref{fig:5_panelD}). 

Interestingly, when comparing phase diagrams with two demixing regions, we find that the tie lines have always a positive gradient -- \emph{i.e.}, the concentration of counterions is lower in the dilute (I) instead of condensed phase (II). In contrast, for the phase diagrams with a single demixing region, we observe different trends in the tie-lines: (e)-(g) always a positive gradient; (h) a mix of tie-lines with positive and negative gradients in the proximity of the critical point. 

Overall, we find that, similarly to fixed charges, the presence of charge-regulating binding sites lowers the demixing tendency of polymers (compared to the neutral case -- see shaded area in~\Cref{fig:6}). Nonetheless, we find that charge regulation mechanisms, unlike fixed charges, yield more complex topologies of the phase diagrams. As investigated in the next section, this gives rise to non-linear dependencies between the polymer solubility as a function of the solution acidity. 

\subsubsection{The impact of counterions on polymer solubility.}

Recent studies have focused on studying how chemical properties of salt ions (such as the counterion radii) affect the solubility of charged polymers with fixed charges~\cite{duan2023_arxiv}. Their theoretical results, for a system of polyelectrolytes in a solvent with salt (i.e. positive and negative mobile ions), show non-monotonic salt concentration dependence where salting-out at low salt concentrations is due to ionic screening. In the high salt concentration regime, the macromolecules remain in the salting-out regime for small ions but change to a salting-in regime for larger ions. They conclude that the solubility at high salt concentrations is determined by the competition between the solvation energy and the (translational) entropy of ions, addressing the intensely discussed problem of salt effects in LLPS of protein solutions, such as re-entrant phase transitions shown experimentally in \cite{krainer2021reentrant,oh_simple_2023}. 

\begin{figure*}[ht]
\centering
    \begin{subfigure}{0.0\textwidth}
    \captionlistentry{}
    \label{fig:salt_panelA}
    \end{subfigure}    
    \begin{subfigure}{0.0\textwidth}
    \captionlistentry{}
    \label{fig:salt_panelB}
    \end{subfigure}   
    \begin{subfigure}{\textwidth}    
    \includegraphics[width=0.9\textwidth]{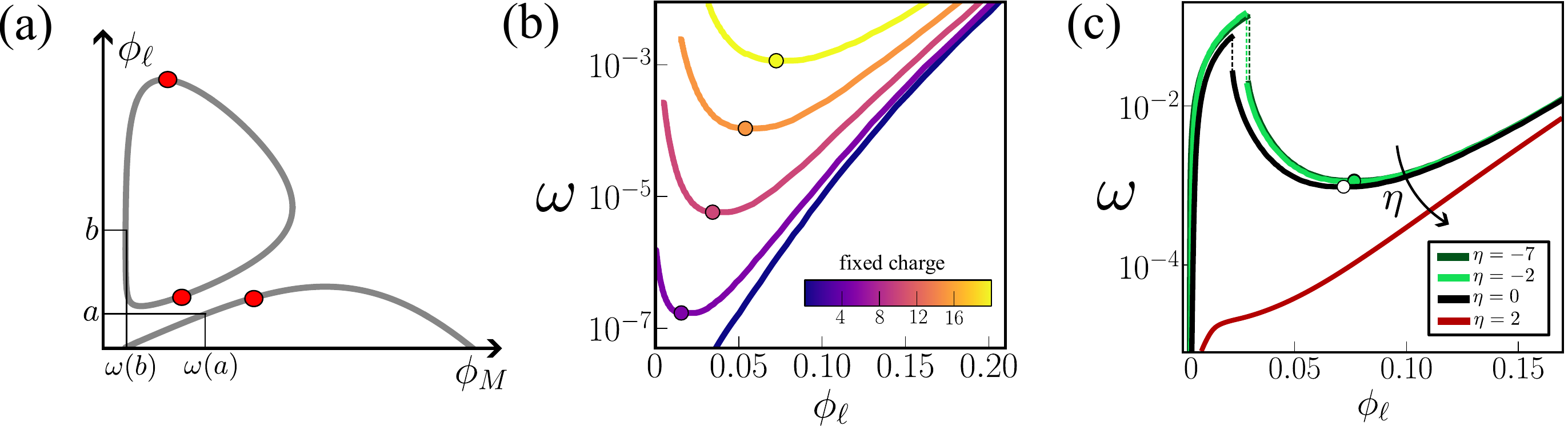}
    \captionlistentry{}
    \label{fig:salt_panelC}
    \end{subfigure}   
    \vspace{-5mm}
    \caption{{\bf Counterion effect on polymer solubility.} (a) Schematic showing how the solubility, $\omega$, is computed starting from the phase diagrams in~\Cref{sec:coexistence_curves} (details in the main text). (b) Solubility $\omega$ as a function of the counterion concentration for polymers with different fixed charges (same parameters as in~\Cref{fig:4a}). (c) Solubility $\omega$ as a function of the counterion concentrations for charge regulating polymers (same parameters as in Figures~\ref{fig:6}).}
    \label{fig:7}
\end{figure*}

Here, we are interested analogously in studying the impact of counterions (or solution acidity) on the solubility of charged polymers. 
We define the solubility, $\omega=\omega(\phi_\ell)$, of a charged polymer for a given counterion concentration $\phi_\ell$, as the minimum value of the equilibrium volume fraction on the binodal curves (see schematic drawing in~\Cref{fig:salt_panelA}). Our definition is analogous to the one used in~\cite{duan2023_arxiv}, but corrected for the fact that, in our model, multiple coexistence curves may exist. 

As shown in~\Cref{fig:salt_panelB}, we find that for neutral molecules the solubility increases with counterion concentration (see purple curve). In contrast, when considering polymers with fixed charge, $\omega$, has a non-monotonic profile which agrees with the results obtained in~\cite{duan2023_arxiv} (for relatively large salt ions), despite our simpler approximation of electrostatic fluctuations. At low counterion concentrations, the solubility of the polymers decreases with $\phi_\ell$. This trend -- which we referred to as conterion-out behaviour -- is considered to be universal for all ions at low ionic concentrations and is explained by the fact that the counterions are able to screen the charge on the polymers and hence reduce the Coulomb repulsion between 
the polymers. In contrast, at higher counterion concentrations, the solubility increases with $\phi_\ell$ -- the conterion-in effect. This can be explained by the dominant contribution of the entropy of mixing associated with the ions over charge-screening effects, which favours the miscibility of the solution, very similar to the properties of the system studied in \cite{duan2023_arxiv}.

As shown in~\Cref{fig:salt_panelC}, solubility curves of charge-regulating polymers present more complex trends. When $\eta\leq 0$, we find that the solubility curve can be split into three regimes: acid-in at extremely low counterion concentrations; acid-out for intermediate-to-low counterion concentrations; counterion-in at high counterion concentrations. Note that in the transition between the counterion-in at extremely low $\phi_\ell$ to counterion-out behaviour for low $\phi_\ell$, the solubility curve is not smooth. Jumps in $\omega$ and $\omega'$ is a signature of the presence and merging of the two disconnected demixing regions (see the curves with $\eta\leq 0$ in~\Cref{fig:salt_panelC}). For larger values of $\eta$ (see red curve in~\Cref{fig:salt_panelC}), corresponding to the scenario where binding of the ions to the monomers is unfavourable, we recover a monotonic solubility curve as for neutral macro-molecules: consistent counterion-in behaviour (independently of $\phi_\ell$).

The transition in the sign of the first derivative  from $\omega'>0$ to $\omega'<0$ is a signature of another important feature of the phase diagrams in Figure~\ref{fig:5_panelA}-\ref{fig:5_panelC}: 
counterion-driven re-entrant phase separation. Specifically, when short-range repulsion are not too strong (see Figures~\ref{fig:5_panelA}-\ref{fig:5_panelC}), the system exhibits re-entrant behaviour when varying the concentration of counterions, $\phi_{\ell}$. In other words, there are values of $\phi_M$ that lie in the demixing region at very low and high values of $\phi_\ell$ but not for intermediate (or very high) concentrations of counterions.

\subsection{Regulation of the charge distribution via phase separation.}
\label{sec:observation_CR}

In the previous section, we have shown how charge regulation affects phase separation in solutions of charged polymers. Conversely, in this section, we are interested in how phase separation itself regulates polymer charge in solution. 
In order to investigate this aspect, we consider a standard quenching experiment where we drive the system to phase separate by controlling the acidity of the solution (\emph{i.e.}, decreasing $\phi_\ell$). Specifically, we start from a homogeneous mixture ($O$) with composition $\phi^{O}_M=0.2$ and $\phi^{O}_\ell=0.04$; this is then perturbed by decreasing the acid volume fraction to $\phi_\ell=0.023$. When considering spatially homogeneous equilibria, at any location in space the charge distribution of the polymer phase is the same. However, this is not guaranteed when considering a demixed solution consisting of a dilute ($\mathrm{I}$) and condensed ($\mathrm{II}$) phase. In this case, we denote by $\pi_{z}^{\mathrm{I}}$ and $\pi_z^{\mathrm{II}}$ the charge of polymers in each of the two phases. When considering the solution as a whole, the charge distribution on the polymers can be expressed as the weighted average of $\pi_{z}^{\mathrm{I}}$ and $\pi_z^{\mathrm{II}}$:
\begin{equation}
\pi_z^{O'}=\frac{\gamma \pi_z^{\mathrm{I}}\phi_M^{\mathrm{I}}+(1-\gamma)\pi_z^{\mathrm{IP}}\phi_M^{\mathrm{II}}}{\gamma\phi_M^{\mathrm{I}}+(1-\gamma)\phi_M^{\mathrm{II}}}
\end{equation}
where $\gamma$ is the fraction of the total volume of the solution occupied by the dilute phase (I) in the quenched state ($O'$). The value of $\gamma$ is constrained by the conservation of the total concentration of any of the species in the solution; without loss of generality we here consider the conservation of the polymer molecules to obtain:
\begin{equation}
    \gamma =\frac{\phi_{M}^{O}-\phi^{\mathrm{II}}_{M}}{\phi_{M}^{\mathrm{I}}-\phi^{\mathrm{II}}_{M}}. 
\end{equation}

\begin{figure*}[htb]
    \begin{subfigure}{\textwidth}
\includegraphics[width=\textwidth]{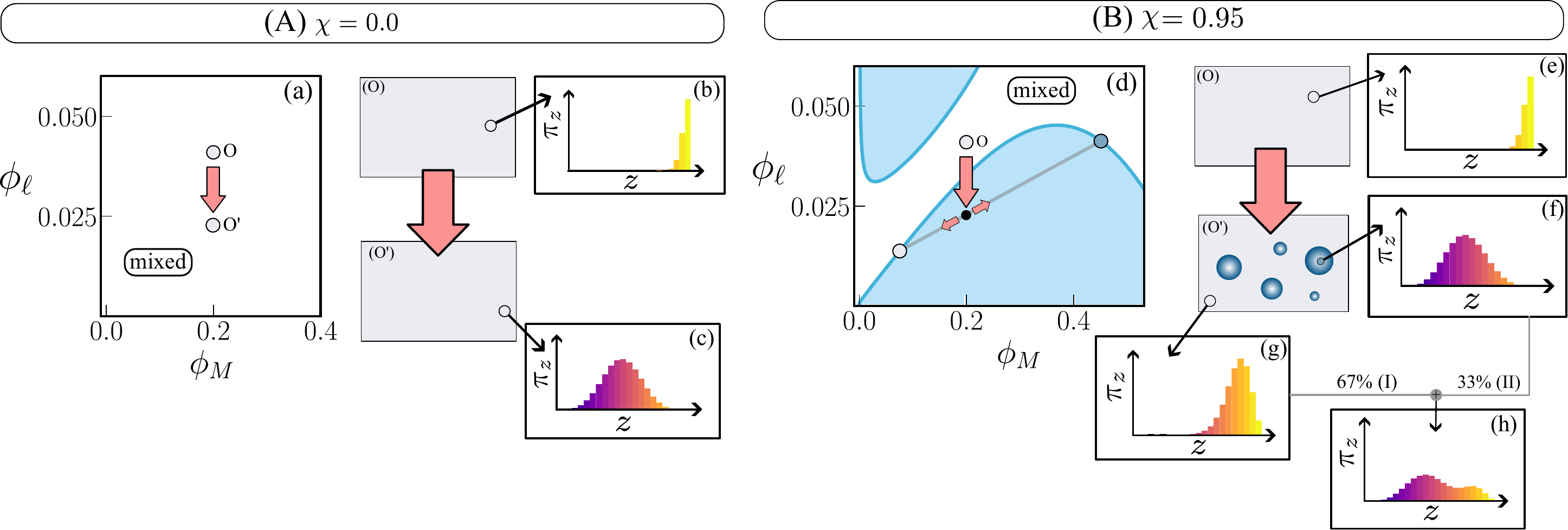}
\captionlistentry{}
\label{fig:8a}
    \end{subfigure}
        \begin{subfigure}{0.0\textwidth}
        \captionlistentry{}
\label{fig:8b}
    \end{subfigure}
        \begin{subfigure}{0.0\textwidth}
        \captionlistentry{}
\label{fig:8c}
    \end{subfigure}
            \begin{subfigure}{0.0\textwidth}
        \captionlistentry{}
\label{fig:8d}
    \end{subfigure}
            \begin{subfigure}{0.0\textwidth}
        \captionlistentry{}
\label{fig:8e}
    \end{subfigure}
            \begin{subfigure}{0.0\textwidth}
        \captionlistentry{}
\label{fig:8f}
    \end{subfigure}
            \begin{subfigure}{0.0\textwidth}
        \captionlistentry{}
\label{fig:8g}
    \end{subfigure}
        \begin{subfigure}{0.0\textwidth}
        \captionlistentry{}
\label{fig:8h}
    \end{subfigure}
    \vspace{-8mm}
    \caption{{\bf Phase separation as a charge regulation mechanism.} Effect of quenching the solution by decreasing the concentration of counterions. We consider two cases: (A) a solution with non-hydrophobic polymers ($\chi=0$); (B) a solution of hydrophobic polymers that phase separates upon quenching ($\chi=0.95$). (a) Phase diagram for the case $\chi=0$ (no demixing). (b) Charge distribution in the initial mixed state ($O$). (c) Charge distribution, $\pi_z$, after the quenching -- \emph{i.e.}, homogeneous mixed state ($O'$). (d) Phase diagram for the case $\chi=0.95$ (same as in~\Cref{fig:5_panelC}). Decreasing the concentration of counterions drives demixing of the solution into a dilute ($\mathrm{I}$) and condensed ($\mathrm{II}$) phase which are determined by the tie lines. The final state ($O'$) is a demixed solution where 67\% of the volume is occupied by the dilute phase while 33\% by the condensed phase. (e) Charge distribution in the initial mixed state ($O$). (f) Local charge distribution for polymer in the condensed phase ($\mathrm{II}$). (g) Local charge distribution for polymer in the condensed phase ($\mathrm{I}$). (h) Overall charge distribution for the demixed mixture ($O'$). Parameter values are set to default values and $\eta=-2$ (as in~\Cref{fig:4}b).}
    \label{fig:8}
\end{figure*}

As a reference case, we test the protocol on a solution of non-hydrophobic polymers that do not phase separate -- see Figure ~\ref{fig:8}A. In this case, decreasing the acid concentration in the solution (\emph{i.e.}, equivalent to decreasing $\phi_\ell$) does not lead to phase separation. Yet, it significantly affects the polymer charge distribution (compare Figures~\ref{fig:8b} and~\ref{fig:8c}), leading to discharging of the polymer binding sites. In~\Cref{fig:8}B, we consider the same ideal protocol applied to a solution with hydrophobic polymers that phase separates in solution when decreasing the acid volume fraction (see~\Cref{fig:8d}). As shown in~\Cref{fig:8e}, the initial charge distribution on the polymers is similar to the one observed on non-hydrophobic polymers (compare with~\Cref{fig:8b}). Upon quenching, the solution phase separates -- state ($O'$) in~\Cref{fig:8}B. Polymers in the dilute phase remain highly charged (see~\Cref{fig:8g}) as in the initial state ($O$), whereas polymers in the condensed phase partially discharge (see~\Cref{fig:8f}) as in the case of non-hydrophobic polymers (see~\Cref{fig:8c}). When considering the overall solution, the different charge distribution in the two phases is reflected in the charge distribution $\pi^{O'}$ having multiple peaks -- see~\Cref{fig:8h}. By controlling the mixture properties locally -- here solution acidity -- phase separation creates two environments: the condensed phase where the charge distribution on polymers is highly sensitive to changes in the solution acidity, and the dilute phase where the charge distribution is robust to the changes in acidity. As a result, phase separation allows spatial confinement of polymers with a specific charge state. Note that, after quenching, in~\Cref{fig:8}A, polymers with intermediate charge appear homogeneously in the solution, while these are only localised in the condensed phase in~\Cref{fig:8}B. The possible functional implication of these findings in the context of biomolecular condensates will be discussed in the next section.

\section{Conclusions and Discussion}

In this work, we considered a minimal model to investigate the interplay of phase separation and charge regulation. For this, we introduced in ~\Cref{sec:model} a system of chargeable polymers, whose charge state is regulated by protonation/ deprotonation processes, in a water-acid solution.

In~\Cref{sec:homogeneous_phase}, we established the homogeneous equilibria states of the system focusing on how the mixture composition -- \emph{i.e.}, the concentration of the polymers, $\phi_M$, and the counterions, $\phi_{\ell}$ -- affects the polymer charge distribution. In doing so, we employed analytical findings which highlighted the key role of the parameter $\eta$, describing bounded charge interactions, in determining the properties of equilibrium charge distribution. 

Our key findings are: 
For $\eta = 0$, the charge distributions in homogeneous states of the system simplified considerably and it can be found to follow a binomial distribution. For $\eta\neq 0$ we showed that by approximating the charge interaction as a continuous function, we can approximate the charge distribution by a Gaussian distribution for a continuous variable in the limit of a large number of charges, as we derive within a saddle-point approximation. Our analysis yielded that for $\eta < -4$, this approximation ceases to be generally valid, as multi-modal distributions can arise depending on the mixture composition.

In~\Cref{sec:coexistence_curves}, we unfolded how charge regulation processes affect phase diagrams of polymer solutions. To do so, we first characterised phase diagrams assuming a fixed charge on the polymers; we then investigated how the topology changes by introducing charge regulation mechanisms. We found that charge regulation processes can affect the phase diagram topology in a nontrivial manner: upon decreasing $\eta$, we observed that the usual demixing region undergoes a change of its topology, in which a closed-loop region
branches off from the original demixing region -- which persists. This contrasts with the phase diagrams of polymers with fixed charges, where at most one demixing region exists. The complex topology of the phase diagram is reflected in the relation between counterions concentration and polymer solubility (in short solubility curves) in an acid-water solution. We find that charge regulation mechanisms have a prominent signature: depending on the charge-interaction parameter $\eta$, due to the re-entrant phase behaviour induced by charge regulation, the solubility curve can exhibit a pronounced jump. This might be a relevant experimental signature for the charge-regulation induced transition in the topology of coexistence curves. 

In addition, our results show that charge regulation has an important impact on the partitioning of counter-ions and thus the gradients of the corresponding tie-lines, which is further complicated by the existence of multi-modal equilibrium states. These findings add to the discussion on salt-partitioning in the complex coacervation of polyelectrolyte ~\cite{Sing2020}. Different theoretical frameworks -- \emph{e.g.}, random phase approximation (RPA) which includes connectivity of the polyelectrolyte, and Liquid-State theories-- have been proposed to explain salt partitioning and tie-lines gradients. Identifying the physical reasons for salt partitioning amongst the increasing number of candidate theories remains an open problem and an active area of research~\cite{Sing2020}. Our results suggest that allowing for charge regulation can also affect tie-lines gradient thus adding yet another layer to this discussion. 

In the last section, ~\Cref{sec:observation_CR}, we investigated the
effect phase separation has on the charge distribution. By discussing an experimental scenario
in which the concentration of counterions in the polymer solution is changed, we demonstrated that phase separation can create local environments with very different charge distributions in the dilute and condensed phases, which in addition are either very similar or quite different from the initial state. Our findings highlight how charge regulation mechanisms can have a significant role in the response of polymer solutions to changes in the physical environment, by introducing a complex coupling between processes occurring at the micro-scale (protonation/deprotonation) and meso-scale (phase separation). Interestingly, similar non-linear effects -- like e.g. re-entrant phase behaviour-- have been recently discussed by Jacobs et al. in a seemingly unrelated system, in which a different molecular process - polymer self-assembly - is discussed in conjunction with phase separation \cite{Li2023_arxiv}. This suggests re-entrant behaviour might be a general feature of systems where phase separation is coupled to a molecular mechanism (such as charge regulation or self-assembly).

On the one hand, charge regulation controls the sensitivity of polymer solutions to environmental changes by allowing for so-called re-entrant demixing behaviour. We expect this non-linear dependence of phase separation on environmental cues to be fundamental in a range of applications to soft matter science, such as in the design of responsive materials, as well as in LLPS of proteins. Salt-induced re-entrant phase separation has been observed for proteins that undergo LLPS in the high-salt regime~\cite{krainer2021reentrant}, including the intensively investigated protein FUS. These observations, together with our and previous theoretical works~\cite{Adame-Arana2020} show the impact of the environment as a driving force of LLPS and adds a further important mechanism to the widely discussed sequence-dependence LLPS of intrinsically disordered proteins.

Conversely, we also find that by affecting the local environment the polymers are in, phase separation itself can regulate the polymer charge state by allowing to spatially confine polymers in a specific charged state -- hence increasing their local volume fraction. This can have important consequences when considering polymers interacting with additional chemical agents, whereby their interactions may be mediated by the polymer charge state. This is the case in the cell cytoplasm. From this point of view, phase separation in cells might function as a regulator of cellular responses to the environment by controlling both the location, as well as the charge state of proteins. 

Despite its simplicity, our model yields a rich and interesting range of behaviours that hint at the importance of charge regulation mechanisms in the formation and properties of condensates. There is therefore scope to extend our theory to investigate whether our findings have relevance to LLPS in cells. This requires extending our model to account for the complexity of biological macromolecules -- such as RNA and proteins. 
For example, in this work, we have assumed the interaction parameter $\chi$ to be independent of the polymer charge state. However, when considering short-range interactions e.g. between polymer chains, these are known to be charge-dependent. 
Therefore, a further natural extension of this work would be to analyse the scenario in which $\chi$ is considered a function of the charge state $z$. This would result in the mixture composition influencing not only the association–dissociation energy parameter, $\alpha$, but also higher-order interactions between binding sites. 

Overall, our results reveal that, even in the simplest system consisting of one polymer species whose charge states undergo a protonation/deprotonation process, the interplay between phase separation and charge regulation mechanisms governs the response of polymer mixtures to environmental changes.


\begin{acknowledgements}
The authors thank Dr Matthew Hennessy and Prof. Sarah Waters for the helpful discussions in the initial phase of this project.
GLC is supported by the UK Engineering and Physical Sciences Research Council (EPSRC), grant number EP/W524335/1.  
\end{acknowledgements}

%
%

\input{appendix}

\bibliography{references}

\end{document}

%% file: appendix.tex
\section{Appendix}

\subsection{Derivation of the chemical potential
condition} 
\label{app:mu}
We consider an incompressible mixture in the ($T$, $V$, $N $)-ensemble with temperature $T$, volume $V$ and particle numbers $N_\omega$-- where $\omega\in\Omega$. At equilibrium, such a system minimises the Helmholtz free energy, $F=F(T,V,\left\{N_\omega\right\}_{\omega\in\Omega})$.  From Euler's relation,
it follows that
\begin{equation}
    F(T,V,\left\{N_\omega\right\}_{\omega \in \Omega})=\sum_{\omega\in\Omega} \mu_\omega N_\omega - pV,\label{eq_app:general_H_free_energy}
\end{equation}
where $p$ is the pressure and $\mu_\omega$ are the chemical potentials of the different components of the mixture. Incompressibility of the mixture implies that the molecular volume $\nu_\omega$ of each component of the mixture is constant; as a result, the volume of the mixture can not be taken as an independent variable but rather as a function of the particles numbers: $V=\sum_{\omega\in\Omega} N_\omega\nu_\omega$. 
Differentiation of $F$ with respect to particle
number leads to the chemical potential condition
\begin{equation}
\mu_\omega=\frac{\partial F}{\partial N_\omega}+p\nu_\omega\,.
\end{equation} 
Transforming now to the Helmholtz free energy density $f(T,\left\{\phi_\omega\right\}_{\omega\in\Omega})=F/V$ with the volume fractions $\phi_\omega=N_\omega\nu_\omega/V$ we obtain, 
performing the necessary differentiations,
\begin{equation}
\begin{aligned}
\mu_\omega&=\nu_\omega f + \nu_\omega\frac{\partial f}{\partial \phi_\omega}
+V\nu_\omega \sum_\sigma \frac{\partial f}{\partial \phi_\sigma} \frac{(-\nu_\sigma N_\sigma)}{V^2}
+\nu_\omega p
\\&=
\nu_\omega \left[p +\left(f 
- \sum_\sigma \frac{\partial f}{\partial \phi_\sigma}\phi_\sigma
\right)\right]
+ \nu_\omega\frac{\partial f}{\partial \phi_\omega}.
\end{aligned}\label{eq_app:general_chemical_potentials}%
\end{equation}

Applying this general relation to our mixture
we find that the chemical potential of the free ions ($\mu_+$, $\mu_\ell$), solvent ($\mu_s$) and $z$-charged polymers ($\mu_z$) are given by

\begin{eqnarray}
\label{cpe}
\mu_+ & = &(p-\Sigma)\nu + k_BT \left[\ln (\phi_+) + 1-\frac{\lambda}{8\pi}\frac{\kappa}{1+\kappa}\right], 
\\
\mu_\ell & = &(p-\Sigma)\nu +k_BT \left[\ln (\phi_\ell) + 1-\frac{\lambda}{8\pi}\frac{\kappa}{1+\kappa}\right], 
\\
\mu_s & = &(p-\Sigma)\nu +k_BT [\ln (\phi_s) + 1+\chi \phi_M],
\\
\nonumber \\
\mu_{z} & = & (p-\Sigma)N\nu 
\\
& +& k_BT\left[u_z+\ln (\phi_z) + 1 
+\chi N \phi_s
-\frac{z\lambda}{8\pi}\frac{\kappa}{1+\kappa}\right]. \nonumber
\end{eqnarray}
The expression for $\Sigma$ arises from
those terms in the round brackets in (\ref{eq_app:general_chemical_potentials}) that
do not cancel out which is only the case for
contributions from $f_2$ and $f_3$. Making use of
the no-void condition for $f_2$ and the electroneutrality condition for $f_3$ one finds
\begin{eqnarray}
    \frac{\nu\Sigma}{k_BT} =\left(1-\phi_M+ \frac{\phi_M}{N}\right)+\chi \phi_M\phi_s + \nonumber \\
    \frac{1}{4\pi}\left(\ln(1+\kappa)-\frac{\kappa}{2}\frac{2+\kappa}{1+\kappa}\right).
    \label{eq:Sigma}
\end{eqnarray}

\subsection{Saddle-point approximation}
\label{app:Gaussian approximation}
When considering $\eta = 0$, the charge distribution, $\left\{\pi_z\right\}_{z=0}^Z$, is binomial. It is well known that a general binomial distribution, $B(Z,p)$, is well-approximated by a Gaussian distribution with the same mean and standard deviation, in the limit $Z\gg 1$ -- provided $p$ is bounded away from its extreme values $0$ and $1$. Here we show that a saddle-point approximation to
the charge distribution is possible provided that $\eta>-4$, guaranteeing that the charge distribution has a unique maximum. 

Substituting the definition of $u_z$ (see~\eqref{eq:u_z}) into~\eqref{eq:charge_distribution}, we obtain that $\pi_z$ reads
\begin{align}
    \pi_z = \frac{\exp\left(-\alpha_{\mbox{\tiny eff}}z-\frac{\eta z^2}{2 Z} + \ln\left[\binom{Z}{z}\right]\right)}{\sum\limits_{k=0}^Z\exp\left(-\alpha_{\mbox{\tiny eff}} k-\frac{\eta k^2}{2 Z} + \ln\left[\binom{Z}{k}\right]\right)}, \quad z=0,\ldots,Z.\label{eq:discrete_charge_distribution} 
\end{align}
where the $\alpha_{\eff}$ is as defined in~\eqref{eq:alpha_eff}. 
In what follows, we want to approximate the distribution~\eqref{eq:discrete_charge_distribution} 
by a Gaussian distribution centred at its mean value $\mathcal{Q}=\sum_{z=0}^Zz\pi_z$ under the assumption that $Z\gg 1$. For our approximation to be valid, the mean charge needs to be sufficiently far from its extreme values, \emph{i.e.}, $\mathcal{Q}\gg0$ and $Z-\mathcal{Q}\gg 0$.  Since $Z\gg1$, we rewrite the discrete charge distribution~(\ref{eq:discrete_charge_distribution}), as a continuous probability distribution for the continuous variable $z\in[0,Z]$. First, we approximate the binomial coefficient by using Stirling’s series:
\begin{equation}
    \ln(n!) \approx n\ln n - n +\frac{1}{2}\ln(2\pi n) + O\left(\frac{1}{n}\right).\label{eq:stirling}
\end{equation}
Using~\eqref{eq:stirling}, we find
\begin{eqnarray}
    \frac{1}{Z} \ln\left[\binom{Z}{z}\right] & = & -\frac{z}{Z}\ln\frac{z}{Z} -\left(1-\frac{z}{Z}\right)\ln\left(1-\frac{z}{Z}\right)
    \nonumber \\
    & - & \frac{1}{2 Z}\ln\left(2\pi z\left(1-\frac{z}{Z}\right)\right) 
    \nonumber \\
    & + & O\left(\frac{1}{Z z}\right)+ O\left(\frac{1}{Z (Z-z)}\right),
\end{eqnarray}
Following the continuous approximation, we can write~\eqref{eq:discrete_charge_distribution} but considering $z\in(0,Z)$ as a continuous distribution:
\begin{subequations}
\begin{align}
    \pi_z = \frac{\exp\left(-Z \tilde{u}_{\mbox{\tiny eff}}\left(\frac{z}{Z}\right)\right)}{Z\int_0^1\exp\left(- Z \tilde{u}_{\mbox{\tiny eff}}(\omega)\right)d\omega},\quad z\in(0,Z), \label{eq:continuous_charge_distribution} 
\end{align}
where
\begin{align}
\begin{aligned}
    \tilde{u}_{\eff}(\omega) = \omega\alpha_{\eff}&+\frac{\omega^2\eta}{2} +\omega\ln\omega \\&+(1-\omega)\ln(1-\omega) + h.o.t.
    \end{aligned}
    \end{align}
\end{subequations}
By computing the second derivative of $\tilde{u}_{\eff}$, it is apparent that for $\eta>-4$ the function $\tilde{u}_{\mbox{\eff}}$ is convex for $\omega\in (0,1)$. This guarantees that there exists a unique minimum, $p\in(0,1)$. As discussed in~\Cref{sec:homogeneous_phase}, we can interpret $p$ as an effective binding probability of ions to the polymer that we have defined in the main text. An implicit definition for $p$ can be obtained by solving $ \tilde{u}_{\eff}' (p)=0$:
\begin{equation}
p = \frac{e^{-\alpha_{\mbox{\tiny eff}}-p\eta}}{1+e^{-\alpha_{\mbox{\tiny eff}}-p\eta}}.\label{eq:stationary}
\end{equation}
When $Z\gg 1$, the mass of the normalisation integral (see first factor in~\Cref{eq:continuous_charge_distribution}) will be localised around the stationary point, $p$, and standard techniques, such as Laplace's method can be applied:
\begin{equation}
    \int_0^1 e^{-Z\tilde{u}_{\eff}(\omega)}d\omega \approx \sqrt{\frac{2\pi}{Z\tilde{u}_{\mbox{\eff}}''(p)}} e^{-Z\tilde{u}_{\eff}(p)},
\end{equation}
where
\begin{subequations}
\begin{align}
    \tilde{u}_{\eff}(p)&=-\frac{\eta p^2}{2}-\ln\left(1+\frac{\phi_+}{\phi_s}e^{-\alpha-\chi \phi_M-p\eta}\right), \\
    \tilde{u}_{\eff}''(p)&=\frac{\eta p(1-p) + 1}{p (1-p)}.
\end{align}
\end{subequations}
Substituting the above into~\Cref{eq:continuous_charge_distribution} and expanding around the stationary point ($Z p$), we obtain a Gaussian distribution:
\begin{align}
    \pi_z \approx \frac{1}{\sqrt{2\pi Z \tilde{u}_{\mbox{\tiny eff}}''(p)}}\exp\left(-\frac{(z-Z p)^{2}}{2 Z \tilde{u}_{\mbox{\tiny eff}}''(p)}\right). \label{eq:continuous_charge_distribution2}
\end{align}

\begin{figure}[htb]
    \centering
    \includegraphics[width=0.475\textwidth]{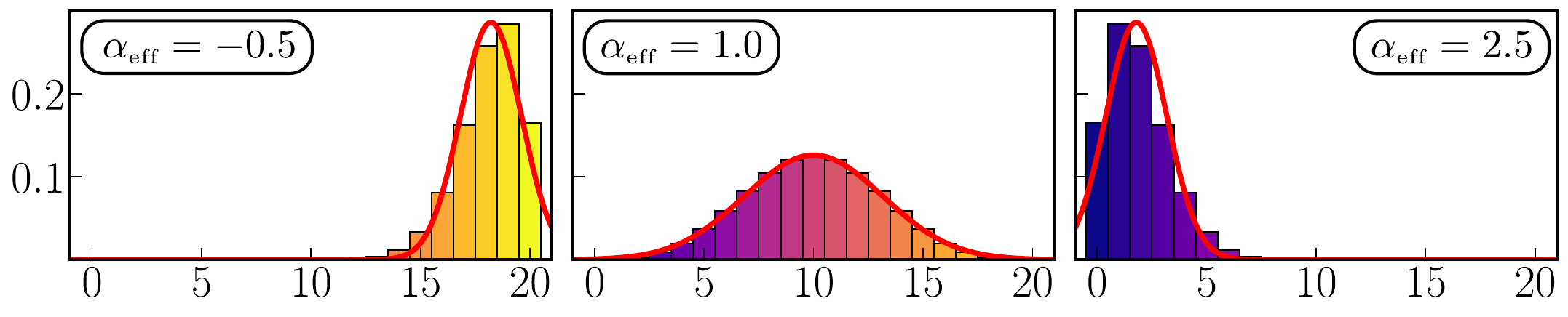}
    \caption{\textbf{Saddle-point approximation.} Plots comparing the exact discrete distribution~\eqref{eq:discrete_charge_distribution} (histogram) and its approximation obtained via the saddle-point approach~\eqref{eq:continuous_charge_distribution2} (red curve). Different panels corresponds to different choices of the parameter $\alpha_{\eff}$.}
    \label{fig:comparison_gauss_approx}
\end{figure}
In~\Cref{fig:comparison_gauss_approx}, we compare the approximated distribution~\eqref{eq:continuous_charge_distribution2} with the real distribution~\eqref{eq:discrete_charge_distribution} for different values of $\alpha_{\eff}$. We find good agreement between the two. Nonetheless, discrepancies emerge when considering $|\alpha_{\eff}|\gg 1$ when the maximum of the distribution shifts towards the boundary of the domain: for $\alpha_{\eff}$ large and negative, the maximum $\approx Z$ while for $\alpha_{\eff}$ large and positive $\approx 0$. This discrepancy is to be expected since for the approximation to hold we must assume $p$ is bounded away the extreme values $0$ and $1$. 

\subsection{Unimodal distribution: domain of physicality.}
\label{app:derivation unimodal}

In this section, we outline results on the existence and uniqueness of the effective binding probability $p$ (see~\eqref{eq_eta_non_zero:implicit_p}) assuming $\eta>-4$ for any values of $\phi_M\in(0,1)$ and $\phi_\ell\in(0,1)$ which are physically allowed.
 
By substituting~\eqref{eq:distribution_unimodal} into~\eqref{eq_intro:no-void}-\eqref{eq_intro:electroneutrality}, we find that $p$ is implicitly defined by the non-linear algebraic equation $\Pi_\eta(p)=0$, where 
\begin{eqnarray}
    \Pi_\eta(x) & = & e^{x \eta}x\left(1-\phi_M-\phi_\ell\right)  \label{def_Pi_p}\\
    & + &\left(\frac{Z \,x}{N} \phi_M-\phi_\ell\right)\left[x e^{\eta x}+e^{-\alpha-\chi\phi_M}(1-x)\right]\, .
    \nonumber 
\end{eqnarray}
The form of $\Pi_\eta$ is obtained starting from ~\eqref{eq:stationary} and~\eqref{eq:alpha_eff} by first eliminating $\phi_+$
via~\eqref{eq_intro:electroneutrality} in the form 
$\phi_+= \phi_\ell - Z p /N \phi_M $, and by finally using ~\eqref{eq_intro:no-void} in the form $\phi_s = 1+(Zp/N-1)\phi_M - 2\phi_{\ell} $ to eliminate the
remaining dependence on $\phi_s$. Note that when setting $\eta=0$, $\Pi_0$ reduces to a quadratic equation for $p_0$ that can be solved explicitly:
 \begin{subequations}
 \begin{align}
     p_{0}=\frac{-b(\phi_M,\phi_\ell)+\sqrt{b^2(\phi_M,\phi_\ell)+4c(\phi_M,\phi_\ell)}}{2},
 \end{align}
where
\begin{align}
    b(\phi_M,\phi_\ell)&=\frac{N}{Z}\frac{1-\phi_M-2\phi_\ell+e^{-\alpha-\chi\phi_M}(Z/N\phi_M+\phi_\ell)}{(1-e^{-\alpha-\chi\phi_M}) \phi_M},\\
    c(\phi_M,\phi_\ell)&=\frac{N}{Z}\frac{\phi_\ell e^{-\alpha-\chi\phi_M}}{(1-e^{-\alpha-\chi\phi_M}) \phi_M}.
\end{align}
 \end{subequations}

Nonetheless, we consider the more general case $\eta>-4$, and prove that there exists at most one root $p$ for the function $\Pi_{\eta}$ in the interval $(0,1]$; conditions for existence are then discussed. We here exclude $0$ since $p=0$ refers to the critical case where no counter-ions are present in the solution: $\Pi_\eta(0)=-e^{-\alpha-\chi\phi_M}\phi_\ell=0$ only if $\phi_\ell=0$. First, we note that, for $x\in (0,1]$, the first term 
on the right hand side in~\eqref{def_Pi_p} is always non-negative (since from the no-void condition $\phi_s+\phi_+=1-\phi_\ell-\phi_M\geq 0$). The sign of the second term instead depends on the value of $(Zx/N)\phi_M-\phi_\ell$. Given that $p$ is a root of $\Pi_\eta$, if that exists, we must have that $Z p/N \phi_M-\phi_\ell<0$ (which guarantees that $\phi_+>0$). 

We now consider the value of the first derivative of $\Pi_\eta$ and evaluate it at one of its possible roots:
\begin{eqnarray}
    \Pi_\eta'(p) & = & \frac{Z}{N} \phi_M\left(p e^{p\eta}+e^{-\alpha-\chi\phi_M}(1-p)\right) 
    \label{eq_app:pi_derivative}\\
    & + & \left(\phi_\ell-\frac{Z p}{N}  \phi_M\right)e^{-\alpha-\chi\phi_M}\frac{1+\eta p(1-p)}{p}, \nonumber 
\end{eqnarray}
where we have used the fact that $\Pi_\eta(p)=0$. It is apparent that the first term in~\eqref{eq_app:pi_derivative} is positive and so is the second term, since, as discussed above, we must have that $\phi_\ell-Z/N p \phi_M>0$. This implies that for any root of $\Pi_\eta$, $p\in(0,1]$, the derivative $\Pi_\eta'(p)>0$. 

\paragraph{Uniqueness.} Since the function $\Pi_\eta$ is analytic and $\Pi_\eta'(p)>0$, we conclude that if $p$ exists this must be unique. Otherwise, there would exists a root $\tilde{p}\in(0,1)$ such that $\Pi_\eta'(\tilde{p})\leq 0$. 

\paragraph{Existence.} Generally, the existence of $p$ is not guaranteed. Since $\Pi_\eta(0)<0$, the only conditions for the existence of $p$ is that $\Pi_\eta(1)\geq 0$:
\begin{equation}
1-\phi_M-2\phi_\ell+\frac{Z}{N}\phi_M>0.
\end{equation}

\paragraph{Domain of physicality.} Summarising the results above, we find two inequality constraints that homogeneous equilibria exist (\emph{i.e}, $p$ is well-defined) provided that:
\begin{subequations}
    \begin{align}
        1-\phi_M-\phi_\ell >0,\label{eq_app:existence_1}\\
        1+\phi_M \left(\frac{Z}{N}-1\right) - 2\phi_\ell>0.\label{eq_app:existence_2}
    \end{align}\label{eq_app:cond}%
\end{subequations}
However, for the equilibria to be physical meaning, we must have that the corresponding volume fractions $\phi_+$ and $\phi_s$ are positive and less than one. Conditions~\eqref{eq_app:cond} are sufficient to guarantee this is the case.

\subsection{Two-phase coexistence conditions}
\label{app:ph}

In this section, we derive the coexistence conditions used to compute the phase diagrams presented in~\Cref{sec:coexistence_curves}. 
We start by considering an initially homogeneous mixture
of the $Z+4$ species that has been quenched into the unstable
regime, just before separates into two phases. Each of the emerging phases are homogeneous with a unique composition, characterised by the composition vectors
$\phi_\omega^I$ and $\phi_\omega^{II}$. 
In the \emph{demixed} state, the conditions for the coexistence of two phases are 
\begin{equation}
\mu(\{\phi^I_\omega\}_{\omega\in\Omega})
=
\mu(\{\phi^{II}_\omega\}_{\omega\in\Omega}).
\end{equation}
These are $Z+4$ conditions for $2(Z+4)$ variables, leaving $Z+4$ degrees of
freedom. For the charge regulation (CR) model, we assume each phase is in chemical equilibrium, which imposes the chemical potentials in each of the two phases to satisfy Eq.~\eqref{app_eq_chem_costraint}, or $Z$ restrictions each. When considering the fixed charge (FC) model, the system is constrained by imposing the charge distribution $\pi_z=\delta(z-Z)$ in both phases; also in the latter case, this leads to $2Z$ restrictions. However, due to the equality of chemical
potentials between phases, we only need to impose these $Z$ conditions on one phase (for the other they are then
implied). So we have 4 degrees of freedom left. We  also have to satisfy
electroneutrality and no-void in each phase, which removes all four remaining
degrees of freedom. As a result, we lack one degree of freedom 
required to match those of the initial homogeneous mixture
prior to demixing.

This problem is frequently addressed by adding an additional
contribution to the chemical potential $\mu_\omega$ in \eqref{cpe} for each of
the charged species, giving rise to the electrochemical 
potential,
\begin{equation}
\check\mu_\omega=\mu_\omega+z_\omega e \psi,
\end{equation}
where $\psi$ is the Galvani potential. These electrochemical
potentials are then equated instead of the chemical potentials.
In a homogeneous system, the Galvani potential $\psi$ is
constant and hence can be eliminated by setting it to zero, but in a
non-homogeneous e.g.\ demixed system it is usually not. 
We then have two different values for $\psi$ and the difference between the
two remains as the previously missing additional degree of freedom.

Here, we proceed differently to motivate the introduction of
a Galvani potential and describe the phase separation as a minimisation problem. 
We again consider a system 
with two coexisting phases and a total
volume $V=1$ (without loss of generality), split into two sub-systems I and
II of volume $\gamma$ and $1-\gamma$, respectively, with $0<\gamma<1$. 
Each subsystem is occupied by a single, in itself homogeneous phase 
described by the variables
$\vec \phi^I =\left(\phi_\omega^I\right)_{\omega\in\Omega}$ and
$\vec\phi^{II}=\left(\phi_\omega^{II}\right)_{\omega\in\Omega}$, respectively. The total free energy
of the demixed system is then given by
\begin{equation}\label{F2}
F_2=\gamma f(\vec \phi^I)
	+(1-\gamma) f(\vec\phi^{II}).
\end{equation}
In a system without chemical reactions, each species is individually subject to mass conservation
and we would minimise $F$ under these $Z+4$ constraints to find the equilibrium of the system.
With chemical reactions, a smaller number of quantities are conserved, and these 
quantities need to be determined in an additional step prior to the formulation of
the minimisation problem. For this purpose, note that the total number of molecules 
of species $\omega$ present is given by
\begin{equation}
N_\omega=\left(\gamma  \phi_\omega^I + (1-\gamma) \phi_\omega^{II}\right)/\nu_\omega
\end{equation} 
For 
\begin{equation}
K=\sum_{\omega\in\Omega} a_\omega  N_\omega
\end{equation} to be conserved, the vector $\vec a=(a_\omega)_{\omega\in\Omega}$
has to satisfy
\begin{equation}\label{sta}
{\mathbb S}^T \vec a=0,
\end{equation}
where $\mathbb S$ is the stoichiometric matrix (with $Z+4$ rows and $Z$ columns), that is, the rows of its transpose are the stoichiometric coefficients of the
chemical reactions. To write out this matrix, we assume that the indices $\omega$
are ordered as $z=0,1,\ldots, Z$ followed by $s,+,\ell$. Then we get
\begin{equation}
{\mathbb S}^T=
\begin{bmatrix}
-1 & 1 & 0 & 0 &\ldots & 1 & -1 & 0 \\
0 & -1 & 1 & 0 & \ldots & 1 & -1 & 0 \\
\vdots & \ddots & \ddots & \ddots & \ddots & \vdots & \vdots & \vdots\\
0 & 0 & \ldots &- 1 & 1 & 1 & -1 & 0
\end{bmatrix}.
\end{equation}
Four linearly independent solutions $\vec a$ of \eqref{sta} can be easily read off and
give the conserved quantities
\begin{subequations}\label{cc}
\begin{align}
K_1&=\nu N_\ell=\gamma  \phi_\ell^I + (1-\gamma) \phi_\ell^{II},\label{cca}\\
\nonumber \\
K_2&=\nu\left(N_s+N_+\right)  = \gamma \left( \phi_s^I+\phi_+^I \right) 
				+ (1-\gamma) \left( \phi_s^{II}+\phi_+^{II} \right), \label{ccb} \nonumber \\
\\    
K_3&=\nu_M\sum_{z=0}^Z N_z = 
\gamma \sum_{z=0}^Z\phi_z^I + (1-\gamma) \sum_{z=0}^Z\phi_z^{II}, \label{ccc} 
\\
K_4&=\nu\left(\sum_{z=0}^Z z N_z+N_+\right) \nonumber \\
&=\gamma \left( \frac{\nu}{\nu_M}\left(\sum_{z=0}^Z z\phi_z^I\right)+\phi_+^I\right) 
\label{ccd}\\
&+(1-\gamma)  \left(\frac{\nu}{\nu_M}\left(\sum_{z=0}^Z z\phi_z^{II}\right)+\phi_+^{II}\right) \nonumber
\end{align}
\end{subequations}
In the minimisation problem for $F_2$ in \eqref{F2}, we enforce that the
$K_i\equiv K_i(\vec\phi^I,\vec\phi^{II},\gamma)$ are equal to a constant
parameter $K^0_i$, the value of which is set for example by the composition of
the mixture prior to separation into two phases. We impose the resulting
conditions as constraints, alongside the no-void \eqref{eq_intro:no-void} and
electroneutrality \eqref{eq_intro:electroneutrality} conditions enforced
separately for each of the two phases. However, it turns out that \eqref{ccc} is
implied by \eqref{cca}, \eqref{ccb} and the no-void condition
\eqref{eq_intro:no-void}, and therefore can be dropped.  Similarly, \eqref{ccd}
is implied by \eqref{cca} and electroneutrality
\eqref{eq_intro:electroneutrality}, so this constraint can be dropped, too.

Including the constraints via Lagrange multipliers $\lambda_1$, $\lambda_2$,
$\rho^I$, $\rho^{II}$ and $\psi^I$, $\psi^{II}$, we seek the stationary points of
\begin{eqnarray}
L_2&=&
F_2+\lambda_1(K_1-K_1^0)+\lambda_2(K_2-K_2^0) \nonumber \\
& +& \rho^I \left(\sum_{\omega\in\Omega} \phi_\omega^I-1\right)
+\rho^{II} \left(\sum_{\omega\in\Omega} \phi_\omega^{II}-1\right) \nonumber \\
& +& \psi^I e\gamma\left(\phi^I_+-\phi^I_\ell+\frac{\nu}{\nu_M}
\sum_{z=1}^{Z} z \phi^I_z\right) \\
& + &\psi^{II} e(1-\gamma) \left(\phi^{II}_+-\phi^{II}_\ell+\frac{\nu}{\nu_M}
\sum_{z=1}^{Z} z \phi^{II}_z\right) \nonumber
\end{eqnarray}
Notice that we have weighted the electroneutrality conditions with
with the elementary charge $e$ and with the relative volume 
$\gamma$ and $1-\gamma$ occupied by phase $I$ and $II$, respectively.

By differentiating $L_2$ with respect to $\phi_+^I$ and $\phi_l^I$, we get
\begin{equation}
\frac{\partial f}{\partial \phi_+^I}
+\lambda_2-\frac{\partial f}{\partial \phi_\ell^I}-\lambda_1+2\psi^I=0,
\end{equation}
and similarly
for phase $II$. Subtracting the expressions for the two phases
and using \eqref{psisum}-\eqref{DH} to evaluate the derivatives of $f$, we obtain
\begin{equation}
 e\frac{\psi^{II}-\psi^I}{k_BT} = 
\frac{1}{2}\ln
\left[\frac{\phi_+^{I}}{\phi_\ell^I} \frac{\phi_\ell^{II}}{\phi_+^{II}}\right], 
\label{eq_equilibrium :maxwell_construction}
\end{equation}
The difference $\psi^{II}-\psi^I$ can be identified with the
net potential jump due to the electric field between the
two phases, also known as \emph{Galvani potential} \cite{Zhang2021}.

Returning to $L_2$ and setting its first derivatives with respect to the 
components of $\vec\phi^I$ to zero, we obtain,
after some algebra, the condition
\begin{eqnarray}
\nu_M\left(\frac{\partial f}{\partial \phi^I_{z}}-\frac{\partial f}{\partial \phi^I_{z-1}}\right)
=
\nu\left(\frac{\partial f}{\partial \phi^I_+}-\frac{\partial f}{\partial \phi^I_s}\right), \\
\nonumber \\
\qquad z=1\ldots Z+1; \nonumber 
\end{eqnarray}
similarly for $I$ replaced by $II$. This is exactly the condition
\eqref{app_eq_chem_costraint}, applied to each phase; see also
\eqref{eq_app:general_chemical_potentials}. We therefore can use 
\eqref{eq:charge_distribution}, together with \eqref{defphiP}, 
to eliminate the $\phi_z^I$ and $\phi_z^{II}$ variables, to get the minimisation problem
\begin{subequations}\label{min4}
\begin{eqnarray}
 \gamma f(\phi^I_s,\phi^I_+,\phi^I_\ell,\phi^I_M)
+ (1-\gamma) f(\phi^{II}_s,\phi^{II}_+,\phi^{II}_\ell,\phi^{II}_M) 
&=& \min!,\nonumber\\
\end{eqnarray}
subject to the constraints
\begin{align}
\gamma  \phi_\ell^I + (1-\gamma) \phi_\ell^{II}&=K_1^0,\label{min4b}\\
\gamma  \phi_M^I + (1-\gamma) \phi_M^{II}&=K_5^0,\label{min4c}\\
\phi^R_s+\phi^R_++\phi^R_\ell+\phi^R_M&=1, \quad R=I,\,II,\label{min4d}\\
\phi^R_+-\phi^R_\ell+\frac{\nu}{\nu_M}\phi^R_M {\cal Q}^R&=0, \quad R=I,\,II,\label{min4e}
\end{align}
with constants $K_1^0$ and $K_5^0$.  Notice that \eqref{min4c}
replaces \eqref{ccb} by a linear combination of the other constraints. 
\end{subequations}

We treat this minimisation problem by using \eqref{min4d} and \eqref{min4e}
to eliminate the $\phi^R_s$ and $\phi^R_+$ variables (for $R=I,\,II$) from $f$
(and denote it by $f^*$)
but including \eqref{min4b} and \eqref{min4c} via Lagrange multipliers. 
Differentiating with respect to $\phi_l^R$, $\phi_M^R$ and $\gamma$ gives the conditions 
\begin{eqnarray}
\label{eqs:tangent_construction1}
\mu^*_M(\phi_M^I,\phi_\ell^I)=\mu^*_M(\phi_M^{II},\phi_\ell^{II}),\\
\mu^*_{\ell}(\phi_M^I,\phi_\ell^I)=\mu^*_{\ell}(\phi_M^{II},\phi_\ell^{II}),\\
\mu^*_M(\phi_M^I-\phi_M^{{II}})+\mu^*_{\ell}(\phi_\ell^I-\phi_\ell^{{II}}) &=& \\
	f^*(\phi_M^I,\phi_\ell^I)-f^*(\phi_M^{II},\phi_\ell^{II}), \nonumber
\end{eqnarray}
where 
\begin{align}
\mu^*_M&=\partial f^*/\partial \phi_M
\label{mustar1}
\intertext{and}
\mu^*_{\ell}&=\partial f^*/\partial \phi_\ell,
\label{mustar2}
\end{align}
These are the equations we solve using bifurcation packages as described in
the main text; $f^*$ is substituted by either $f_{\mbox{\tiny CR}}$ (see~\eqref{eq:reformulated_free_energy}) or $f_{\mbox{\tiny FC}}$ (see~\eqref{eq:f_FC}) depending on the formulation of the model of interest.

%% file: jcp_bcmw.bbl
\begin{thebibliography}{39}%
\makeatletter
\providecommand \@ifxundefined [1]{%
 \@ifx{#1\undefined}
}%
\providecommand \@ifnum [1]{%
 \ifnum #1\expandafter \@firstoftwo
 \else \expandafter \@secondoftwo
 \fi
}%
\providecommand \@ifx [1]{%
 \ifx #1\expandafter \@firstoftwo
 \else \expandafter \@secondoftwo
 \fi
}%
\providecommand \natexlab [1]{#1}%
\providecommand \enquote  [1]{``#1''}%
\providecommand \bibnamefont  [1]{#1}%
\providecommand \bibfnamefont [1]{#1}%
\providecommand \citenamefont [1]{#1}%
\providecommand \href@noop [0]{\@secondoftwo}%
\providecommand \href [0]{\begingroup \@sanitize@url \@href}%
\providecommand \@href[1]{\@@startlink{#1}\@@href}%
\providecommand \@@href[1]{\endgroup#1\@@endlink}%
\providecommand \@sanitize@url [0]{\catcode `\\12\catcode `\$12\catcode
  `\&12\catcode `\#12\catcode `\^12\catcode `\_12\catcode `\%12\relax}%
\providecommand \@@startlink[1]{}%
\providecommand \@@endlink[0]{}%
\providecommand \url  [0]{\begingroup\@sanitize@url \@url }%
\providecommand \@url [1]{\endgroup\@href {#1}{\urlprefix }}%
\providecommand \urlprefix  [0]{URL }%
\providecommand \Eprint [0]{\href }%
\providecommand \doibase [0]{https://doi.org/}%
\providecommand \selectlanguage [0]{\@gobble}%
\providecommand \bibinfo  [0]{\@secondoftwo}%
\providecommand \bibfield  [0]{\@secondoftwo}%
\providecommand \translation [1]{[#1]}%
\providecommand \BibitemOpen [0]{}%
\providecommand \bibitemStop [0]{}%
\providecommand \bibitemNoStop [0]{.\EOS\space}%
\providecommand \EOS [0]{\spacefactor3000\relax}%
\providecommand \BibitemShut  [1]{\csname bibitem#1\endcsname}%
\let\auto@bib@innerbib\@empty
\bibitem [{\citenamefont {Riback}\ \emph {et~al.}(2020)\citenamefont {Riback},
  \citenamefont {Zhu}, \citenamefont {Ferrolino}, \citenamefont {Tolbert},
  \citenamefont {Mitrea}, \citenamefont {Sanders}, \citenamefont {Wei},
  \citenamefont {Kriwacki},\ and\ \citenamefont {Brangwynne}}]{Riback2020}%
  \BibitemOpen
  \bibfield  {author} {\bibinfo {author} {\bibfnamefont {J.~A.}\ \bibnamefont
  {Riback}}, \bibinfo {author} {\bibfnamefont {L.}~\bibnamefont {Zhu}},
  \bibinfo {author} {\bibfnamefont {M.~C.}\ \bibnamefont {Ferrolino}}, \bibinfo
  {author} {\bibfnamefont {M.}~\bibnamefont {Tolbert}}, \bibinfo {author}
  {\bibfnamefont {D.~M.}\ \bibnamefont {Mitrea}}, \bibinfo {author}
  {\bibfnamefont {D.~W.}\ \bibnamefont {Sanders}}, \bibinfo {author}
  {\bibfnamefont {M.-T.}\ \bibnamefont {Wei}}, \bibinfo {author} {\bibfnamefont
  {R.~W.}\ \bibnamefont {Kriwacki}},\ and\ \bibinfo {author} {\bibfnamefont
  {C.~P.}\ \bibnamefont {Brangwynne}},\ }\bibfield  {title} {\bibinfo {title}
  {Composition-dependent thermodynamics of intracellular phase separation},\
  }\href {https://doi.org/10.1038/s41586-020-2256-2} {\bibfield  {journal}
  {\bibinfo  {journal} {Nature}\ }\textbf {\bibinfo {volume} {581}},\ \bibinfo
  {pages} {1476} (\bibinfo {year} {2020})}\BibitemShut {NoStop}%
\bibitem [{\citenamefont {Villegas}\ \emph {et~al.}(2022)\citenamefont
  {Villegas}, \citenamefont {Heidenreich},\ and\ \citenamefont
  {Levy}}]{Villegas2022}%
  \BibitemOpen
  \bibfield  {author} {\bibinfo {author} {\bibfnamefont {J.}~\bibnamefont
  {Villegas}}, \bibinfo {author} {\bibfnamefont {M.}~\bibnamefont
  {Heidenreich}},\ and\ \bibinfo {author} {\bibfnamefont {E.~D.}\ \bibnamefont
  {Levy}},\ }\bibfield  {title} {\bibinfo {title} {Molecular and environmental
  determinants of biomolecular condensate formation},\ }\href
  {https://doi.org/10.1038/s41589-022-01175-4} {\bibfield  {journal} {\bibinfo
  {journal} {Nature Chemical Biology}\ }\textbf {\bibinfo {volume} {18}},\
  \bibinfo {pages} {1319} (\bibinfo {year} {2022})}\BibitemShut {NoStop}%
\bibitem [{\citenamefont {Choi}\ \emph {et~al.}(2020)\citenamefont {Choi},
  \citenamefont {Holehouse},\ and\ \citenamefont {Pappu}}]{choi_physical_2020}%
  \BibitemOpen
  \bibfield  {author} {\bibinfo {author} {\bibfnamefont {J.-M.}\ \bibnamefont
  {Choi}}, \bibinfo {author} {\bibfnamefont {A.~S.}\ \bibnamefont
  {Holehouse}},\ and\ \bibinfo {author} {\bibfnamefont {R.~V.}\ \bibnamefont
  {Pappu}},\ }\bibfield  {title} {\bibinfo {title} {Physical principles
  underlying the complex biology of intracellular phase transitions},\ }\href
  {https://doi.org/10.1146/annurev-biophys-121219-081629} {\bibfield  {journal}
  {\bibinfo  {journal} {Annual Review of Biophysics}\ }\textbf {\bibinfo
  {volume} {49}},\ \bibinfo {pages} {107} (\bibinfo {year} {2020})}\BibitemShut
  {NoStop}%
\bibitem [{\citenamefont {Shapiro}\ \emph {et~al.}(2021)\citenamefont
  {Shapiro}, \citenamefont {Ney}, \citenamefont {Eghtesadi},\ and\
  \citenamefont {Chilkoti}}]{shapiro_protein_2021}%
  \BibitemOpen
  \bibfield  {author} {\bibinfo {author} {\bibfnamefont {D.~M.}\ \bibnamefont
  {Shapiro}}, \bibinfo {author} {\bibfnamefont {M.}~\bibnamefont {Ney}},
  \bibinfo {author} {\bibfnamefont {S.~A.}\ \bibnamefont {Eghtesadi}},\ and\
  \bibinfo {author} {\bibfnamefont {A.}~\bibnamefont {Chilkoti}},\ }\bibfield
  {title} {\bibinfo {title} {Protein phase separation arising from intrinsic
  disorder: First-principles to bespoke applications},\ }\href
  {https://doi.org/10.1021/acs.jpcb.1c01146} {\bibfield  {journal} {\bibinfo
  {journal} {The Journal of Physical Chemistry B}\ }\textbf {\bibinfo {volume}
  {125}},\ \bibinfo {pages} {6740} (\bibinfo {year} {2021})}\BibitemShut
  {NoStop}%
\bibitem [{\citenamefont {Overbeek}\ and\ \citenamefont
  {Voorn}(1957)}]{VO_theory}%
  \BibitemOpen
  \bibfield  {author} {\bibinfo {author} {\bibfnamefont {J.~T.~G.}\
  \bibnamefont {Overbeek}}\ and\ \bibinfo {author} {\bibfnamefont {M.~J.}\
  \bibnamefont {Voorn}},\ }\bibfield  {title} {\bibinfo {title} {Phase
  separation in polyelectrolyte solutions. theory of complex coacervation},\
  }\href {https://doi.org/https://doi.org/10.1002/jcp.1030490404} {\bibfield
  {journal} {\bibinfo  {journal} {Journal of Cellular and Comparative
  Physiology}\ }\textbf {\bibinfo {volume} {49}},\ \bibinfo {pages} {7}
  (\bibinfo {year} {1957})}\BibitemShut {NoStop}%
\bibitem [{\citenamefont {Nott}\ \emph {et~al.}(2015)\citenamefont {Nott},
  \citenamefont {Petsalaki}, \citenamefont {Farber}, \citenamefont {Jervis},
  \citenamefont {Fussner}, \citenamefont {Plochowietz}, \citenamefont {Craggs},
  \citenamefont {Bazett-Jones}, \citenamefont {Pawson}, \citenamefont
  {Forman-Kay} \emph {et~al.}}]{nott2015phase}%
  \BibitemOpen
  \bibfield  {author} {\bibinfo {author} {\bibfnamefont {T.~J.}\ \bibnamefont
  {Nott}}, \bibinfo {author} {\bibfnamefont {E.}~\bibnamefont {Petsalaki}},
  \bibinfo {author} {\bibfnamefont {P.}~\bibnamefont {Farber}}, \bibinfo
  {author} {\bibfnamefont {D.}~\bibnamefont {Jervis}}, \bibinfo {author}
  {\bibfnamefont {E.}~\bibnamefont {Fussner}}, \bibinfo {author} {\bibfnamefont
  {A.}~\bibnamefont {Plochowietz}}, \bibinfo {author} {\bibfnamefont {T.~D.}\
  \bibnamefont {Craggs}}, \bibinfo {author} {\bibfnamefont {D.~P.}\
  \bibnamefont {Bazett-Jones}}, \bibinfo {author} {\bibfnamefont
  {T.}~\bibnamefont {Pawson}}, \bibinfo {author} {\bibfnamefont {J.~D.}\
  \bibnamefont {Forman-Kay}}, \emph {et~al.},\ }\bibfield  {title} {\bibinfo
  {title} {Phase transition of a disordered nuage protein generates
  environmentally responsive membraneless organelles},\ }\href
  {https://doi.org/10.1016/j.molcel.2015.01.013} {\bibfield  {journal}
  {\bibinfo  {journal} {Molecular cell}\ }\textbf {\bibinfo {volume} {57}},\
  \bibinfo {pages} {936} (\bibinfo {year} {2015})}\BibitemShut {NoStop}%
\bibitem [{\citenamefont {Lin}\ \emph {et~al.}(2016)\citenamefont {Lin},
  \citenamefont {Forman-Kay},\ and\ \citenamefont {Chan}}]{Lin2016}%
  \BibitemOpen
  \bibfield  {author} {\bibinfo {author} {\bibfnamefont {Y.-H.}\ \bibnamefont
  {Lin}}, \bibinfo {author} {\bibfnamefont {J.~D.}\ \bibnamefont
  {Forman-Kay}},\ and\ \bibinfo {author} {\bibfnamefont {H.~S.}\ \bibnamefont
  {Chan}},\ }\bibfield  {title} {\bibinfo {title} {Sequence-specific
  polyampholyte phase separation in membraneless organelles},\ }\href
  {https://doi.org/10.1103/PhysRevLett.117.178101} {\bibfield  {journal}
  {\bibinfo  {journal} {Phys. Rev. Lett.}\ }\textbf {\bibinfo {volume} {117}},\
  \bibinfo {pages} {178101} (\bibinfo {year} {2016})}\BibitemShut {NoStop}%
\bibitem [{\citenamefont {Meca}\ \emph {et~al.}(2023)\citenamefont {Meca},
  \citenamefont {Fritsch}, \citenamefont {Iglesias-Artola}, \citenamefont
  {Reber},\ and\ \citenamefont {Wagner}}]{Meca2023}%
  \BibitemOpen
  \bibfield  {author} {\bibinfo {author} {\bibfnamefont {E.}~\bibnamefont
  {Meca}}, \bibinfo {author} {\bibfnamefont {A.~W.}\ \bibnamefont {Fritsch}},
  \bibinfo {author} {\bibfnamefont {J.~M.}\ \bibnamefont {Iglesias-Artola}},
  \bibinfo {author} {\bibfnamefont {S.}~\bibnamefont {Reber}},\ and\ \bibinfo
  {author} {\bibfnamefont {B.}~\bibnamefont {Wagner}},\ }\bibfield  {title}
  {\bibinfo {title} {Predicting disordered regions driving phase separation of
  proteins under variable salt concentration},\ }\bibfield  {journal} {\bibinfo
   {journal} {Frontiers in Physics}\ }\textbf {\bibinfo {volume} {11}},\ \href
  {https://doi.org/10.3389/fphy.2023.1213304} {10.3389/fphy.2023.1213304}
  (\bibinfo {year} {2023})\BibitemShut {NoStop}%
\bibitem [{\citenamefont {Zhang}\ \emph {et~al.}(2020)\citenamefont {Zhang},
  \citenamefont {Vigers}, \citenamefont {McCarty}, \citenamefont {Rauch},
  \citenamefont {Fredrickson}, \citenamefont {Wilson}, \citenamefont {Shea},
  \citenamefont {Han},\ and\ \citenamefont {Kosik}}]{zhang2020proline}%
  \BibitemOpen
  \bibfield  {author} {\bibinfo {author} {\bibfnamefont {X.}~\bibnamefont
  {Zhang}}, \bibinfo {author} {\bibfnamefont {M.}~\bibnamefont {Vigers}},
  \bibinfo {author} {\bibfnamefont {J.}~\bibnamefont {McCarty}}, \bibinfo
  {author} {\bibfnamefont {J.~N.}\ \bibnamefont {Rauch}}, \bibinfo {author}
  {\bibfnamefont {G.~H.}\ \bibnamefont {Fredrickson}}, \bibinfo {author}
  {\bibfnamefont {M.~Z.}\ \bibnamefont {Wilson}}, \bibinfo {author}
  {\bibfnamefont {J.-E.}\ \bibnamefont {Shea}}, \bibinfo {author}
  {\bibfnamefont {S.}~\bibnamefont {Han}},\ and\ \bibinfo {author}
  {\bibfnamefont {K.~S.}\ \bibnamefont {Kosik}},\ }\bibfield  {title} {\bibinfo
  {title} {The proline-rich domain promotes tau liquid--liquid phase separation
  in cells},\ }\href {https://doi.org/10.1083/jcb.202006054} {\bibfield
  {journal} {\bibinfo  {journal} {Journal of Cell Biology}\ }\textbf {\bibinfo
  {volume} {219}},\ \bibinfo {pages} {e202006054} (\bibinfo {year}
  {2020})}\BibitemShut {NoStop}%
\bibitem [{\citenamefont {Brangwynne}\ \emph {et~al.}(2015)\citenamefont
  {Brangwynne}, \citenamefont {Tompa},\ and\ \citenamefont
  {Pappu}}]{Brangwynne2015}%
  \BibitemOpen
  \bibfield  {author} {\bibinfo {author} {\bibfnamefont {C.~P.}\ \bibnamefont
  {Brangwynne}}, \bibinfo {author} {\bibfnamefont {P.}~\bibnamefont {Tompa}},\
  and\ \bibinfo {author} {\bibfnamefont {R.~V.}\ \bibnamefont {Pappu}},\
  }\bibfield  {title} {\bibinfo {title} {Polymer physics of intracellular phase
  transitions},\ }\href {https://doi.org/10.1038/nphys3532} {\bibfield
  {journal} {\bibinfo  {journal} {Nature Physics}\ }\textbf {\bibinfo {volume}
  {11}},\ \bibinfo {pages} {899} (\bibinfo {year} {2015})}\BibitemShut
  {NoStop}%
\bibitem [{\citenamefont {Sing}(2017)}]{Sing20172}%
  \BibitemOpen
  \bibfield  {author} {\bibinfo {author} {\bibfnamefont {C.~E.}\ \bibnamefont
  {Sing}},\ }\bibfield  {title} {\bibinfo {title} {Development of the modern
  theory of polymeric complex coacervation},\ }\href
  {https://doi.org/https://doi.org/10.1016/j.cis.2016.04.004} {\bibfield
  {journal} {\bibinfo  {journal} {Advances in Colloid and Interface Science}\
  }\textbf {\bibinfo {volume} {239}},\ \bibinfo {pages} {2} (\bibinfo {year}
  {2017})},\ \bibinfo {note} {complex Coacervation: Principles and
  Applications}\BibitemShut {NoStop}%
\bibitem [{\citenamefont {Sing}\ and\ \citenamefont {Perry}(2020)}]{Sing2020}%
  \BibitemOpen
  \bibfield  {author} {\bibinfo {author} {\bibfnamefont {C.~E.}\ \bibnamefont
  {Sing}}\ and\ \bibinfo {author} {\bibfnamefont {S.~L.}\ \bibnamefont
  {Perry}},\ }\bibfield  {title} {\bibinfo {title} {Recent progress in the
  science of complex coacervation},\ }\href
  {https://doi.org/10.1039/D0SM00001A} {\bibfield  {journal} {\bibinfo
  {journal} {Soft Matter}\ }\textbf {\bibinfo {volume} {16}},\ \bibinfo {pages}
  {2885} (\bibinfo {year} {2020})}\BibitemShut {NoStop}%
\bibitem [{\citenamefont {Rumyantsev}\ \emph {et~al.}(2021)\citenamefont
  {Rumyantsev}, \citenamefont {Jackson},\ and\ \citenamefont
  {De~Pablo}}]{rumyantsev2021polyelectrolyte}%
  \BibitemOpen
  \bibfield  {author} {\bibinfo {author} {\bibfnamefont {A.~M.}\ \bibnamefont
  {Rumyantsev}}, \bibinfo {author} {\bibfnamefont {N.~E.}\ \bibnamefont
  {Jackson}},\ and\ \bibinfo {author} {\bibfnamefont {J.~J.}\ \bibnamefont
  {De~Pablo}},\ }\bibfield  {title} {\bibinfo {title} {Polyelectrolyte complex
  coacervates: Recent developments and new frontiers},\ }\href
  {https://doi.org/10.1146/annurev-conmatphys-042020-113457} {\bibfield
  {journal} {\bibinfo  {journal} {Annual Review of Condensed Matter Physics}\
  }\textbf {\bibinfo {volume} {12}},\ \bibinfo {pages} {155} (\bibinfo {year}
  {2021})}\BibitemShut {NoStop}%
\bibitem [{\citenamefont {Englander}\ \emph {et~al.}(1997)\citenamefont
  {Englander}, \citenamefont {Mayne}, \citenamefont {Bai},\ and\ \citenamefont
  {Sosnick}}]{englander_hydrogen_1997}%
  \BibitemOpen
  \bibfield  {author} {\bibinfo {author} {\bibfnamefont {S.~W.}\ \bibnamefont
  {Englander}}, \bibinfo {author} {\bibfnamefont {L.}~\bibnamefont {Mayne}},
  \bibinfo {author} {\bibfnamefont {Y.}~\bibnamefont {Bai}},\ and\ \bibinfo
  {author} {\bibfnamefont {T.~R.}\ \bibnamefont {Sosnick}},\ }\bibfield
  {title} {\bibinfo {title} {Hydrogen exchange: The modern legacy of
  {L}inderstrøm-{L}ang},\ }\href@noop {} {\bibfield  {journal} {\bibinfo
  {journal} {Protein Science}\ }\textbf {\bibinfo {volume} {6}},\ \bibinfo
  {pages} {1101} (\bibinfo {year} {1997})}\BibitemShut {NoStop}%
\bibitem [{\citenamefont {Pace}\ \emph {et~al.}(2009)\citenamefont {Pace},
  \citenamefont {Grimsley},\ and\ \citenamefont {Scholtz}}]{pace_protein_2009}%
  \BibitemOpen
  \bibfield  {author} {\bibinfo {author} {\bibfnamefont {C.~N.}\ \bibnamefont
  {Pace}}, \bibinfo {author} {\bibfnamefont {G.~R.}\ \bibnamefont {Grimsley}},\
  and\ \bibinfo {author} {\bibfnamefont {J.~M.}\ \bibnamefont {Scholtz}},\
  }\bibfield  {title} {\bibinfo {title} {{Protein ionizable groups: pK values
  and their contribution to protein stability and solubility}},\ }\href
  {https://doi.org/10.1074/jbc.R800080200} {\bibfield  {journal} {\bibinfo
  {journal} {Journal of Biological Chemistry}\ }\textbf {\bibinfo {volume}
  {284}},\ \bibinfo {pages} {13285} (\bibinfo {year} {2009})}\BibitemShut
  {NoStop}%
\bibitem [{\citenamefont {Tanford}\ and\ \citenamefont
  {Kirkwood}(1957)}]{tanford_theory_1957}%
  \BibitemOpen
  \bibfield  {author} {\bibinfo {author} {\bibfnamefont {C.}~\bibnamefont
  {Tanford}}\ and\ \bibinfo {author} {\bibfnamefont {J.~G.}\ \bibnamefont
  {Kirkwood}},\ }\bibfield  {title} {\bibinfo {title} {Theory of protein
  titration curves. i. general equations for impenetrable spheres},\
  }\href@noop {} {\bibfield  {journal} {\bibinfo  {journal} {Journal of the
  American Chemical Society}\ }\textbf {\bibinfo {volume} {79}},\ \bibinfo
  {pages} {5333} (\bibinfo {year} {1957})}\BibitemShut {NoStop}%
\bibitem [{\citenamefont {Avni}\ \emph {et~al.}(2019)\citenamefont {Avni},
  \citenamefont {Andelman},\ and\ \citenamefont {Podgornik}}]{Avni2019}%
  \BibitemOpen
  \bibfield  {author} {\bibinfo {author} {\bibfnamefont {Y.}~\bibnamefont
  {Avni}}, \bibinfo {author} {\bibfnamefont {D.}~\bibnamefont {Andelman}},\
  and\ \bibinfo {author} {\bibfnamefont {R.}~\bibnamefont {Podgornik}},\
  }\bibfield  {title} {\bibinfo {title} {Charge regulation with fixed and
  mobile charged macromolecules},\ }\href
  {https://doi.org/https://doi.org/10.1016/j.coelec.2018.10.014} {\bibfield
  {journal} {\bibinfo  {journal} {Current Opinion in Electrochemistry}\
  }\textbf {\bibinfo {volume} {13}},\ \bibinfo {pages} {70} (\bibinfo {year}
  {2019})}\BibitemShut {NoStop}%
\bibitem [{\citenamefont {Avni}\ \emph {et~al.}(2020)\citenamefont {Avni},
  \citenamefont {Podgornik},\ and\ \citenamefont {Andelman}}]{Avni2020}%
  \BibitemOpen
  \bibfield  {author} {\bibinfo {author} {\bibfnamefont {Y.}~\bibnamefont
  {Avni}}, \bibinfo {author} {\bibfnamefont {R.}~\bibnamefont {Podgornik}},\
  and\ \bibinfo {author} {\bibfnamefont {D.}~\bibnamefont {Andelman}},\
  }\bibfield  {title} {\bibinfo {title} {Critical behavior of charge-regulated
  macro-ions},\ }\href {https://doi.org/10.1063/5.0011623} {\bibfield
  {journal} {\bibinfo  {journal} {The Journal of Chemical Physics}\ }\textbf
  {\bibinfo {volume} {153}},\ \bibinfo {pages} {024901} (\bibinfo {year}
  {2020})}\BibitemShut {NoStop}%
\bibitem [{\citenamefont {Adame-Arana}\ \emph {et~al.}(2020)\citenamefont
  {Adame-Arana}, \citenamefont {Weber}, \citenamefont {Zaburdaev},
  \citenamefont {Prost},\ and\ \citenamefont
  {J{\"{u}}licher}}]{Adame-Arana2020}%
  \BibitemOpen
  \bibfield  {author} {\bibinfo {author} {\bibfnamefont {O.}~\bibnamefont
  {Adame-Arana}}, \bibinfo {author} {\bibfnamefont {C.~A.}\ \bibnamefont
  {Weber}}, \bibinfo {author} {\bibfnamefont {V.}~\bibnamefont {Zaburdaev}},
  \bibinfo {author} {\bibfnamefont {J.}~\bibnamefont {Prost}},\ and\ \bibinfo
  {author} {\bibfnamefont {F.}~\bibnamefont {J{\"{u}}licher}},\ }\bibfield
  {title} {\bibinfo {title} {{Liquid Phase Separation Controlled by pH}},\
  }\href {https://doi.org/10.1016/J.BPJ.2020.07.044} {\bibfield  {journal}
  {\bibinfo  {journal} {Biophysical Journal}\ }\textbf {\bibinfo {volume}
  {119}},\ \bibinfo {pages} {1590} (\bibinfo {year} {2020})}\BibitemShut
  {NoStop}%
\bibitem [{\citenamefont {Muthukumar}\ \emph {et~al.}(2010)\citenamefont
  {Muthukumar}, \citenamefont {Hua},\ and\ \citenamefont
  {Kundagrami}}]{Muthukumar2010}%
  \BibitemOpen
  \bibfield  {author} {\bibinfo {author} {\bibfnamefont {M.}~\bibnamefont
  {Muthukumar}}, \bibinfo {author} {\bibfnamefont {J.}~\bibnamefont {Hua}},\
  and\ \bibinfo {author} {\bibfnamefont {A.}~\bibnamefont {Kundagrami}},\
  }\bibfield  {title} {\bibinfo {title} {Charge regularization in phase
  separating polyelectrolyte solutions},\ }\href
  {https://doi.org/10.1063/1.3328821} {\bibfield  {journal} {\bibinfo
  {journal} {The Journal of Chemical Physics}\ }\textbf {\bibinfo {volume}
  {132}},\ \bibinfo {pages} {084901} (\bibinfo {year} {2010})}\BibitemShut
  {NoStop}%
\bibitem [{\citenamefont {Hua}\ \emph {et~al.}(2012)\citenamefont {Hua},
  \citenamefont {Mitra},\ and\ \citenamefont {Muthukumar}}]{Jing2012}%
  \BibitemOpen
  \bibfield  {author} {\bibinfo {author} {\bibfnamefont {J.}~\bibnamefont
  {Hua}}, \bibinfo {author} {\bibfnamefont {M.~K.}\ \bibnamefont {Mitra}},\
  and\ \bibinfo {author} {\bibfnamefont {M.}~\bibnamefont {Muthukumar}},\
  }\bibfield  {title} {\bibinfo {title} {Theory of volume transition in
  polyelectrolyte gels with charge regularization},\ }\href
  {https://doi.org/10.1063/1.3698168} {\bibfield  {journal} {\bibinfo
  {journal} {The Journal of Chemical Physics}\ }\textbf {\bibinfo {volume}
  {136}},\ \bibinfo {pages} {134901} (\bibinfo {year} {2012})}\BibitemShut
  {NoStop}%
\bibitem [{\citenamefont {Salehi}\ and\ \citenamefont
  {Larson}(2016)}]{Salehi2016}%
  \BibitemOpen
  \bibfield  {author} {\bibinfo {author} {\bibfnamefont {A.}~\bibnamefont
  {Salehi}}\ and\ \bibinfo {author} {\bibfnamefont {R.~G.}\ \bibnamefont
  {Larson}},\ }\bibfield  {title} {\bibinfo {title} {A molecular thermodynamic
  model of complexation in mixtures of oppositely charged polyelectrolytes with
  explicit account of charge association/dissociation},\ }\href
  {https://doi.org/10.1021/acs.macromol.6b01464} {\bibfield  {journal}
  {\bibinfo  {journal} {Macromolecules}\ }\textbf {\bibinfo {volume} {49}},\
  \bibinfo {pages} {9706} (\bibinfo {year} {2016})},\ \Eprint
  {https://arxiv.org/abs/https://doi.org/10.1021/acs.macromol.6b01464}
  {https://doi.org/10.1021/acs.macromol.6b01464} \BibitemShut {NoStop}%
\bibitem [{\citenamefont {da~Silva}\ \emph {et~al.}(2018)\citenamefont
  {da~Silva}, \citenamefont {Derreumaux},\ and\ \citenamefont
  {Pasquali}}]{da2018protein}%
  \BibitemOpen
  \bibfield  {author} {\bibinfo {author} {\bibfnamefont {F.~L.~B.}\
  \bibnamefont {da~Silva}}, \bibinfo {author} {\bibfnamefont {P.}~\bibnamefont
  {Derreumaux}},\ and\ \bibinfo {author} {\bibfnamefont {S.}~\bibnamefont
  {Pasquali}},\ }\bibfield  {title} {\bibinfo {title} {Protein-{RNA}
  complexation driven by the charge regulation mechanism},\ }\href@noop {}
  {\bibfield  {journal} {\bibinfo  {journal} {Biochemical and biophysical
  research communications}\ }\textbf {\bibinfo {volume} {498}},\ \bibinfo
  {pages} {264} (\bibinfo {year} {2018})}\BibitemShut {NoStop}%
\bibitem [{\citenamefont {Nap}\ \emph {et~al.}(2022)\citenamefont {Nap},
  \citenamefont {Qiao}, \citenamefont {LC}, \citenamefont {Stupp},
  \citenamefont {Olvera de~la Cruz},\ and\ \citenamefont {Szleifer}}]{Nap2016}%
  \BibitemOpen
  \bibfield  {author} {\bibinfo {author} {\bibfnamefont {R.}~\bibnamefont
  {Nap}}, \bibinfo {author} {\bibfnamefont {B.}~\bibnamefont {Qiao}}, \bibinfo
  {author} {\bibfnamefont {P.}~\bibnamefont {LC}}, \bibinfo {author}
  {\bibfnamefont {S.}~\bibnamefont {Stupp}}, \bibinfo {author} {\bibfnamefont
  {M.}~\bibnamefont {Olvera de~la Cruz}},\ and\ \bibinfo {author}
  {\bibfnamefont {I.}~\bibnamefont {Szleifer}},\ }\bibfield  {title} {\bibinfo
  {title} {Acid-base equilibrium and dielectric environment regulate charge in
  supramolecular nanofibers},\ }\bibfield  {journal} {\bibinfo  {journal}
  {Front Chem.}\ }\href {https://doi.org/10.3389/fchem.2022.852164}
  {10.3389/fchem.2022.852164} (\bibinfo {year} {2022})\BibitemShut {NoStop}%
\bibitem [{\citenamefont {Zheng}\ \emph {et~al.}(2021)\citenamefont {Zheng},
  \citenamefont {Avni}, \citenamefont {Andelman},\ and\ \citenamefont
  {Podgornik}}]{Zheng2021}%
  \BibitemOpen
  \bibfield  {author} {\bibinfo {author} {\bibfnamefont {B.}~\bibnamefont
  {Zheng}}, \bibinfo {author} {\bibfnamefont {Y.}~\bibnamefont {Avni}},
  \bibinfo {author} {\bibfnamefont {D.}~\bibnamefont {Andelman}},\ and\
  \bibinfo {author} {\bibfnamefont {R.}~\bibnamefont {Podgornik}},\ }\bibfield
  {title} {\bibinfo {title} {{Phase Separation of Polyelectrolytes: The Effect
  of Charge Regulation}},\ }\href {https://doi.org/10.1021/ACS.JPCB.1C01986}
  {\bibfield  {journal} {\bibinfo  {journal} {The Journal of Physical Chemistry
  B}\ ,\ \bibinfo {pages} {acs.jpcb.1c01986}} (\bibinfo {year}
  {2021})}\BibitemShut {NoStop}%
\bibitem [{\citenamefont {Yekymov}\ \emph {et~al.}(2023)\citenamefont
  {Yekymov}, \citenamefont {Attia}, \citenamefont {Levi-Kalisman},
  \citenamefont {Bitton},\ and\ \citenamefont
  {Yerushalmi-Rozen}}]{Yekymov2023}%
  \BibitemOpen
  \bibfield  {author} {\bibinfo {author} {\bibfnamefont {E.}~\bibnamefont
  {Yekymov}}, \bibinfo {author} {\bibfnamefont {D.}~\bibnamefont {Attia}},
  \bibinfo {author} {\bibfnamefont {Y.}~\bibnamefont {Levi-Kalisman}}, \bibinfo
  {author} {\bibfnamefont {R.}~\bibnamefont {Bitton}},\ and\ \bibinfo {author}
  {\bibfnamefont {R.}~\bibnamefont {Yerushalmi-Rozen}},\ }\bibfield  {title}
  {\bibinfo {title} {Charge regulation of poly(acrylic acid) in solutions of
  non-charged polymer and colloids},\ }\bibfield  {journal} {\bibinfo
  {journal} {Polymers}\ }\textbf {\bibinfo {volume} {15}},\ \href
  {https://doi.org/10.3390/polym15051121} {10.3390/polym15051121} (\bibinfo
  {year} {2023})\BibitemShut {NoStop}%
\bibitem [{\citenamefont {Levin}(2002)}]{YanLevin_2002}%
  \BibitemOpen
  \bibfield  {author} {\bibinfo {author} {\bibfnamefont {Y.}~\bibnamefont
  {Levin}},\ }\bibfield  {title} {\bibinfo {title} {Electrostatic correlations:
  from plasma to biology},\ }\href
  {https://doi.org/10.1088/0034-4885/65/11/201} {\bibfield  {journal} {\bibinfo
   {journal} {Reports on Progress in Physics}\ }\textbf {\bibinfo {volume}
  {65}},\ \bibinfo {pages} {1577} (\bibinfo {year} {2002})}\BibitemShut
  {NoStop}%
\bibitem [{\citenamefont {Qin}\ and\ \citenamefont {de~Pablo}(2016)}]{Qin2016}%
  \BibitemOpen
  \bibfield  {author} {\bibinfo {author} {\bibfnamefont {J.}~\bibnamefont
  {Qin}}\ and\ \bibinfo {author} {\bibfnamefont {J.~J.}\ \bibnamefont
  {de~Pablo}},\ }\bibfield  {title} {\bibinfo {title} {Criticality and
  connectivity in macromolecular charge complexation},\ }\href
  {https://doi.org/https://doi.org/10.1021/acs.macromol.6b02113} {\bibfield
  {journal} {\bibinfo  {journal} {Macromolecules}\ }\textbf {\bibinfo {volume}
  {49}},\ \bibinfo {pages} {8789} (\bibinfo {year} {2016})}\BibitemShut
  {NoStop}%
\bibitem [{\citenamefont {Shen}\ and\ \citenamefont {Wang}(2017)}]{Shen2017}%
  \BibitemOpen
  \bibfield  {author} {\bibinfo {author} {\bibfnamefont {K.}~\bibnamefont
  {Shen}}\ and\ \bibinfo {author} {\bibfnamefont {Z.-G.}\ \bibnamefont
  {Wang}},\ }\bibfield  {title} {\bibinfo {title} {{Electrostatic correlations
  and the polyelectrolyte self energy}},\ }\bibfield  {journal} {\bibinfo
  {journal} {The Journal of Chemical Physics}\ }\textbf {\bibinfo {volume}
  {146}},\ \href {https://doi.org/10.1063/1.4975777} {10.1063/1.4975777}
  (\bibinfo {year} {2017}),\ \bibinfo {note} {084901}\BibitemShut {NoStop}%
\bibitem [{\citenamefont {Friedowitz}\ \emph {et~al.}(2018)\citenamefont
  {Friedowitz}, \citenamefont {Salehi}, \citenamefont {Larson},\ and\
  \citenamefont {Qin}}]{Friedowitz2018}%
  \BibitemOpen
  \bibfield  {author} {\bibinfo {author} {\bibfnamefont {S.}~\bibnamefont
  {Friedowitz}}, \bibinfo {author} {\bibfnamefont {A.}~\bibnamefont {Salehi}},
  \bibinfo {author} {\bibfnamefont {R.~G.}\ \bibnamefont {Larson}},\ and\
  \bibinfo {author} {\bibfnamefont {J.}~\bibnamefont {Qin}},\ }\bibfield
  {title} {\bibinfo {title} {{Role of electrostatic correlations in
  polyelectrolyte charge association}},\ }\bibfield  {journal} {\bibinfo
  {journal} {The Journal of Chemical Physics}\ }\textbf {\bibinfo {volume}
  {149}},\ \href {https://doi.org/10.1063/1.5034454} {10.1063/1.5034454}
  (\bibinfo {year} {2018}),\ \bibinfo {note} {163335}\BibitemShut {NoStop}%
\bibitem [{\citenamefont {Zhang}\ \emph {et~al.}(2018)\citenamefont {Zhang},
  \citenamefont {Alsaifi}, \citenamefont {Wu},\ and\ \citenamefont
  {Wang}}]{zhang_polyelectrolyte_2018}%
  \BibitemOpen
  \bibfield  {author} {\bibinfo {author} {\bibfnamefont {P.}~\bibnamefont
  {Zhang}}, \bibinfo {author} {\bibfnamefont {N.~M.}\ \bibnamefont {Alsaifi}},
  \bibinfo {author} {\bibfnamefont {J.}~\bibnamefont {Wu}},\ and\ \bibinfo
  {author} {\bibfnamefont {Z.-G.}\ \bibnamefont {Wang}},\ }\bibfield  {title}
  {\bibinfo {title} {Polyelectrolyte complex coacervation: {Effects} of
  concentration asymmetry},\ }\href {https://doi.org/10.1063/1.5028524}
  {\bibfield  {journal} {\bibinfo  {journal} {J. Chem. Phys.}\ }\textbf
  {\bibinfo {volume} {149}},\ \bibinfo {pages} {163303} (\bibinfo {year}
  {2018})}\BibitemShut {NoStop}%
\bibitem [{\citenamefont {Zhang}\ and\ \citenamefont {Wang}(2021)}]{Zhang2021}%
  \BibitemOpen
  \bibfield  {author} {\bibinfo {author} {\bibfnamefont {P.}~\bibnamefont
  {Zhang}}\ and\ \bibinfo {author} {\bibfnamefont {Z.-G.}\ \bibnamefont
  {Wang}},\ }\bibfield  {title} {\bibinfo {title} {Interfacial structure and
  tension of polyelectrolyte complex coacervates},\ }\href
  {https://doi.org/10.1021/acs.macromol.1c01809} {\bibfield  {journal}
  {\bibinfo  {journal} {Macromolecules}\ }\textbf {\bibinfo {volume} {54}},\
  \bibinfo {pages} {10994} (\bibinfo {year} {2021})}\BibitemShut {NoStop}%
\bibitem [{\citenamefont {Kumari}\ \emph {et~al.}(2022)\citenamefont {Kumari},
  \citenamefont {Dwivedi},\ and\ \citenamefont {Podgornik}}]{Kumari2022}%
  \BibitemOpen
  \bibfield  {author} {\bibinfo {author} {\bibfnamefont {S.}~\bibnamefont
  {Kumari}}, \bibinfo {author} {\bibfnamefont {S.}~\bibnamefont {Dwivedi}},\
  and\ \bibinfo {author} {\bibfnamefont {R.}~\bibnamefont {Podgornik}},\
  }\bibfield  {title} {\bibinfo {title} {{On the nature of screening in
  Voorn–Overbeek type theories}},\ }\bibfield  {journal} {\bibinfo  {journal}
  {The Journal of Chemical Physics}\ }\textbf {\bibinfo {volume} {156}},\ \href
  {https://doi.org/10.1063/5.0091721} {10.1063/5.0091721} (\bibinfo {year}
  {2022}),\ \bibinfo {note} {244901}\BibitemShut {NoStop}%
\bibitem [{\citenamefont {Fossat}\ \emph {et~al.}(2021)\citenamefont {Fossat},
  \citenamefont {Posey},\ and\ \citenamefont {Pappu}}]{Fossat2021}%
  \BibitemOpen
  \bibfield  {author} {\bibinfo {author} {\bibfnamefont {M.~J.}\ \bibnamefont
  {Fossat}}, \bibinfo {author} {\bibfnamefont {A.~E.}\ \bibnamefont {Posey}},\
  and\ \bibinfo {author} {\bibfnamefont {R.~V.}\ \bibnamefont {Pappu}},\
  }\bibfield  {title} {\bibinfo {title} {Quantifying charge state heterogeneity
  for proteins with multiple ionizable residues},\ }\href
  {https://doi.org/10.1016/j.bpj.2021.11.2886} {\bibfield  {journal} {\bibinfo
  {journal} {Biophysical Journal}\ }\textbf {\bibinfo {volume} {120}},\
  \bibinfo {pages} {5438} (\bibinfo {year} {2021})}\BibitemShut {NoStop}%
\bibitem [{\citenamefont {Veltz}(2020)}]{veltz_bif_kit}%
  \BibitemOpen
  \bibfield  {author} {\bibinfo {author} {\bibfnamefont {R.}~\bibnamefont
  {Veltz}},\ }\href {https://hal.archives-ouvertes.fr/hal-02902346} {\bibinfo
  {title} {{BifurcationKit.jl}}} (\bibinfo {year} {2020})\BibitemShut {NoStop}%
\bibitem [{\citenamefont {Duan}\ and\ \citenamefont
  {Wang}(2023)}]{duan2023_arxiv}%
  \BibitemOpen
  \bibfield  {author} {\bibinfo {author} {\bibfnamefont {C.}~\bibnamefont
  {Duan}}\ and\ \bibinfo {author} {\bibfnamefont {R.}~\bibnamefont {Wang}},\
  }\href@noop {} {\bibinfo {title} {Understanding the salt effects on the
  liquid-liquid phase separation of proteins}} (\bibinfo {year} {2023}),\
  \Eprint {https://arxiv.org/abs/2305.03109} {arXiv:2305.03109} \BibitemShut
  {NoStop}%
\bibitem [{\citenamefont {Krainer}\ \emph {et~al.}(2021)\citenamefont
  {Krainer}, \citenamefont {Welsh}, \citenamefont {Joseph}, \citenamefont
  {Espinosa}, \citenamefont {Wittmann}, \citenamefont {de~Csill{\'e}ry},
  \citenamefont {Sridhar}, \citenamefont {Toprakcioglu}, \citenamefont
  {Gudi{\v{s}}kyt{\.e}}, \citenamefont {Czekalska} \emph
  {et~al.}}]{krainer2021reentrant}%
  \BibitemOpen
  \bibfield  {author} {\bibinfo {author} {\bibfnamefont {G.}~\bibnamefont
  {Krainer}}, \bibinfo {author} {\bibfnamefont {T.~J.}\ \bibnamefont {Welsh}},
  \bibinfo {author} {\bibfnamefont {J.~A.}\ \bibnamefont {Joseph}}, \bibinfo
  {author} {\bibfnamefont {J.~R.}\ \bibnamefont {Espinosa}}, \bibinfo {author}
  {\bibfnamefont {S.}~\bibnamefont {Wittmann}}, \bibinfo {author}
  {\bibfnamefont {E.}~\bibnamefont {de~Csill{\'e}ry}}, \bibinfo {author}
  {\bibfnamefont {A.}~\bibnamefont {Sridhar}}, \bibinfo {author} {\bibfnamefont
  {Z.}~\bibnamefont {Toprakcioglu}}, \bibinfo {author} {\bibfnamefont
  {G.}~\bibnamefont {Gudi{\v{s}}kyt{\.e}}}, \bibinfo {author} {\bibfnamefont
  {M.~A.}\ \bibnamefont {Czekalska}}, \emph {et~al.},\ }\bibfield  {title}
  {\bibinfo {title} {Reentrant liquid condensate phase of proteins is
  stabilized by hydrophobic and non-ionic interactions},\ }\href
  {https://doi.org/10.1038/s41467-021-21181-9} {\bibfield  {journal} {\bibinfo
  {journal} {Nature communications}\ }\textbf {\bibinfo {volume} {12}},\
  \bibinfo {pages} {1085} (\bibinfo {year} {2021})}\BibitemShut {NoStop}%
\bibitem [{\citenamefont {Oh}\ \emph {et~al.}(2023)\citenamefont {Oh},
  \citenamefont {Lee}, \citenamefont {Lee}, \citenamefont {Kim}, \citenamefont
  {Lee}, \citenamefont {Lee},\ and\ \citenamefont {Choi}}]{oh_simple_2023}%
  \BibitemOpen
  \bibfield  {author} {\bibinfo {author} {\bibfnamefont {S.-H.}\ \bibnamefont
  {Oh}}, \bibinfo {author} {\bibfnamefont {J.}~\bibnamefont {Lee}}, \bibinfo
  {author} {\bibfnamefont {M.}~\bibnamefont {Lee}}, \bibinfo {author}
  {\bibfnamefont {S.}~\bibnamefont {Kim}}, \bibinfo {author} {\bibfnamefont
  {W.~B.}\ \bibnamefont {Lee}}, \bibinfo {author} {\bibfnamefont {D.~W.}\
  \bibnamefont {Lee}},\ and\ \bibinfo {author} {\bibfnamefont {S.-H.}\
  \bibnamefont {Choi}},\ }\bibfield  {title} {\bibinfo {title} {Simple
  coacervation of guanidinium-containing polymers induced by monovalent salt},\
  }\href {https://doi.org/10.1021/acs.macromol.2c02346} {\bibfield  {journal}
  {\bibinfo  {journal} {Macromolecules}\ }\textbf {\bibinfo {volume} {56}},\
  \bibinfo {pages} {3989} (\bibinfo {year} {2023})}\BibitemShut {NoStop}%
\bibitem [{\citenamefont {Li}\ \emph {et~al.}(2023)\citenamefont {Li},
  \citenamefont {Rogers},\ and\ \citenamefont {Jacobs}}]{Li2023_arxiv}%
  \BibitemOpen
  \bibfield  {author} {\bibinfo {author} {\bibfnamefont {T.}~\bibnamefont
  {Li}}, \bibinfo {author} {\bibfnamefont {B.}~\bibnamefont {Rogers}},\ and\
  \bibinfo {author} {\bibfnamefont {W.~M.}\ \bibnamefont {Jacobs}},\ }\bibfield
   {title} {\bibinfo {title} {Interplay between self-assembly and phase
  separation in a polymer-complex model},\ }\href@noop {} {\bibfield  {journal}
  {\bibinfo  {journal} {arXiv:2306.13198}\ } (\bibinfo {year}
  {2023})}\BibitemShut {NoStop}%
\end{thebibliography}%
